\documentclass[reprint,amssymb,aps,pra,longbibliography]{revtex4-1}

\usepackage[utf8]{inputenc} 
\usepackage{graphicx} 
\usepackage{xcolor}  
\usepackage{amsmath}
\usepackage{amssymb}	
\usepackage{amstext}

\DeclareSymbolFont{extraup}{U}{zavm}{m}{n}
\DeclareMathSymbol{\varheart}{\mathalpha}{extraup}{86}
\usepackage{textcomp}
\usepackage{bm}
\usepackage{amsthm}
\newtheorem{lemma}{Lemma}

\usepackage{subcaption}

\allowdisplaybreaks

\newcommand{\phasediff}{\Delta\vartheta}

\newcommand{\sysvec}{\alpha}
\newcommand{\memvec}{\alpha}
\newcommand{\fpvec}{\vec{\alpha}^*}
\newcommand{\fpvecindex}{\alpha^*}

\newcommand{\nfalse}{{n^f}}

\newcommand{\cmatrix}{S}
\newcommand{\deltheta}{\Delta\vartheta}

\newcommand{\patterncoordinates}{ pattern coordinates }

\newcommand{\IF}{\,\Leftrightarrow\,}

\newcommand{\sinAv}[1]{\sin{\big(#1\big)}}
\newcommand{\cosAv}[1]{\cos{\big(#1\big)}}
\newcommand{\myvspace}{ }

\newcommand{\matr}[1]{\mathbf{#1}} 

\newcommand{\vspan}{\operatorname{span}}
 
\newcommand{\unstableSet}{q}
\newcommand{\isolatedSet}{p}

\newcommand{\cf}{coupling modulation}

\newcommand{\summneqn}{\sum\limits_{\begin{subarray}{l}	
										m=1\\
										m\neq m'
										\end{subarray}}^M}
										
\newcommand{\newtext}{\textcolor{black}}

\newcommand{\newertext}{\textcolor{black}}
									
\newcommand{\ubar}[1]{\text{\b{$#1$}}}

\newcommand{\limSymbol}{\beta}																	
\begin{document}
\title{A robust autoassociative memory with coupled networks of Kuramoto-type oscillators}
%
\author{Daniel Heger}%
\author{Katharina Krischer}
\email[Contact: ]{daniel.heger@ph.tum.de and krischer@ph.tum.de}
\affiliation{Physics Department, Technical University of Munich, 85748 Garching, James-Franck-Str. 1, Germany}
%
\begin{abstract}
Uncertain recognition success, unfavorable scaling of connection complexity or 
dependence on complex external input impair the usefulness of current 
oscillatory neural networks 
 for pattern recognition or restrict technical realizations to small networks.
We propose a new network architecture of coupled oscillators 
for pattern recognition which shows none of the mentioned flaws.
Furthermore we illustrate the recognition process with simulation 
 results and analyze the new dynamics analytically: Possible output patterns
 are isolated attractors of the system. Additionally, simple criteria for recognition
  success are derived from a lower bound on the basins of attraction.
\end{abstract}
\maketitle
\section{Introduction}
Synchronization of oscillators, i.e., ''the adjustment of rhythms due to 
an interaction'' \cite{pikovsky2003SynchronizationUniversalConcept}, is a ubiquitous concept 
\newertext{representing one mechanism leading to 
collective dynamics}. Its occurrence spreads over 
all scientific disciplines, with applications in engineering, physics, 
chemistry, biology, medicine and even in social sciences. The study of 
synchronization behavior has been correspondingly intense during the 
last two to three decades, the state of the art being summarized in 
recent textbooks, monographs and focus issues 
\cite{pikovsky2003SynchronizationUniversalConcept,
osipov2007SynchronizationOscillatoryNetworks,manrubia2004EmergenceOfDynamicalOrder,
boccaletti2008SynchronizedDynamics,stefanski2009DeterminingThresholds,
nishikawa2011ChaosFocusSynchronization,suykens2008ChaosFocusSynchronization,
kurths2003ChaosFocusSynchronization}.

A particularly intriguing area where synchronization often occurs
 is neuroscience.
The synchronization of neural oscillators controls vital functions but 
is responsible for neural diseases as well. Synchronization phenomena 
are also involved in cognition tasks of the brain 
\cite{hoppensteadt2012WeaklyConnectedNeuralNetworks}. 
The wish to understand and mimic information processing of the brain led 
to a separate field called computational neuroscience. Concomitantly,
 novel types of hardware were proposed that mimic some 
aspects of neural information processing. Their massively parallel 
operation is inherently different from the operating modes 
of all types of processors in everyday hardware.
In this manuscript, we discuss a novel coupling scheme for  
oscillators that generates synchronization patterns
 and thus can be used 
 as an 
autoassociative memory.

When an \emph{autoassociative memory} is 
presented with a defective and/or incomplete piece of data, 
it recognizes and retrieves the correct data from a set of correct 
candidates. From a different point of view, the defective input data are 
mapped onto the most similar of the candidates. 
The ability to ''map'' is also found in complex physical systems:
The trajectory of a
system state will converge to an attractor.
 If several attractors exist, different
sets of initial conditions, called \emph{basins of attraction}, will end up on different attractors.
Therefore, the system ''maps'' all initial conditions within one basin onto its attractor.
Note that few physical systems are actually 
suitable as autoassociative memories: 
First, suitable mappings of the defective data onto the initial conditions and 
from the attractors back onto the correct patterns have to be found. 
Additionally, initial conditions as well as attractors of a system need to be controlled, 
with the latter usually being difficult.
Finally, the initial defective data should be mapped onto the most similar correct data candidate, 
which requires that the basins of attraction actually 
conform with a sensible definition of similarity.
\newertext{T}he idea to use basins of attractions for pattern recognition has originally been proposed by 
 Hopfield for use in neural networks \cite{Hopfield1982_original}. 
 Contributions from mathematics, physics and neuroscience 
 (see the end of \cite{Acebron2005_Kuramoto_report} for a summary) 
 made it possible to merge his ideas with the studies of coupled nonlinear oscillators.

Networks of nonlinear oscillators have been shown to act as autoassociative 
memory devices for binary patterns
 \cite{aonishi1998_phase_transitions,aoyagi_1997_cutting_connections,arenas_1994_longterm,Hoppensteadt1999,Hoelzel2011,Kostorz2013,nishikawa2004_second_order,follmann2014_third_order}
 according to the above-mentioned principle. 
In the \newertext{original} architecture \cite{aonishi1998_phase_transitions,aoyagi_1997_cutting_connections,arenas_1994_longterm}, 
identical Kuramoto oscillators \cite{Acebron2005_Kuramoto_report} 
are fully interconnected via programmable 
connections that can change sign and strength of the coupling according to the 
Hebbian Rule \cite{hebb2005reprint}. If the dynamics are expressed in phase shifts, 
fixed points are the only type of attractors, 
and defective input patterns as well as correct pattern 
candidates can be mapped on two synchronized groups of oscillators whose phases differ 
by $\pi$.

However, this design has two disadvantages:
\begin{itemize}
	\item No distinct, well-separated fixed points exist for the 
	memorized patterns \cite{Hoelzel2015_stability}. Instead, there is 
	one global attractor consisting 
	 of lines of attractive fixed points with 
	neutrally stable eigendirections that connect
	 every memorized pattern with every other. 
	 On short timescales, pattern recognition still works: 
	 Starting at the defective pattern, 
	  the system state quickly relaxes onto the global attractor 
	 close to the most similar pattern. 
	 On the attractor, however, perturbations due to 
	 external noise or implementation inaccuracies dominate and the 
	 system state drifts away from the correct pattern on longer 
	 timescales.
	 Additionally, recognition success cannot be guaranteed as no
	  well-defined basin of attraction exists for any single output 
	  pattern.
	   
	  \item The number of connections scales quadratically with the 
	  number of oscillators, so no large networks can be 
	  implemented in hardware. 
\end{itemize}
So far, no architecture that solves both issues has been proposed. 
However, separate solutions for each problem have been discussed: 
Nishikawa et al. \cite{nishikawa2004_second_order} showed
 that the degeneracy of the attractor can be lifted 
 by adding second order Fourier modes to the coupling. A similar 
 network with third order Fourier modes has been proposed as well 
 \cite{follmann2014_third_order}.
 A partial solution for the scaling problem has been proposed 
  by Hoppensteadt and Izhikevich \cite{Hoppensteadt1999}
 and has been further advanced 
 by Hölzel and Krischer \cite{Hoelzel2011} 
 and Kostorz et al. \cite{Kostorz2013}: Oscillators
  of different frequencies are coupled to the same global coupling  
 that affects every oscillator differently.
 These architectures require an external input of complex time-dependent functions, 
 but the number of connections scales with $\mathcal{O}(N)$. 

\newertext{Here, we propose an architecture that} combines isolated attractors and minimal scaling of 
connection complexity without the need for complex external input.
To this end, we built on previous 
studies \cite{Hoppensteadt1999,Hoelzel2011} of globally 
coupled oscillatory devices, but we introduce
 two peculiar features: Different temporal modulation of the coupling 
 strength and a replacement of the single network by two interconnected 
 subnetworks.  The result is a robust 
autoassociative memory that is straightforward to be implemented as 
hardware and can be readily read out.  Additionally, we can predict 
recognition success analytically. Thus, by exploiting the mutual interaction 
of two sub-networks, we arrive at a network architecture with superior 
functionality.

In the next section, the structure of our new architecture is 
described in detail. Afterwards, we obtain the effective dynamics 
through averaging, analyze 
its fixed points and derive the error-free capacity
  in Sec. \ref{sec:fixed_points}. In order to predict 
recognition success, Sec. \ref{sec:boa_a_matching_crit} derives a 
lower bound on the basins of attraction. The resulting criterion for 
guaranteed matching is then validated with simulations of the full dynamics 
in Sec. \ref{sec:sim_details}. Before results are summarized in
 Sec. \ref{sec:Summary+Outlook}, Sec. \ref{sec:discussion} compares 
 the architecture to previous autoassociative networks of oscillators.
\section{A new scalable Architecture}
\label{sec:architecture}
\begin{figure}
\includegraphics{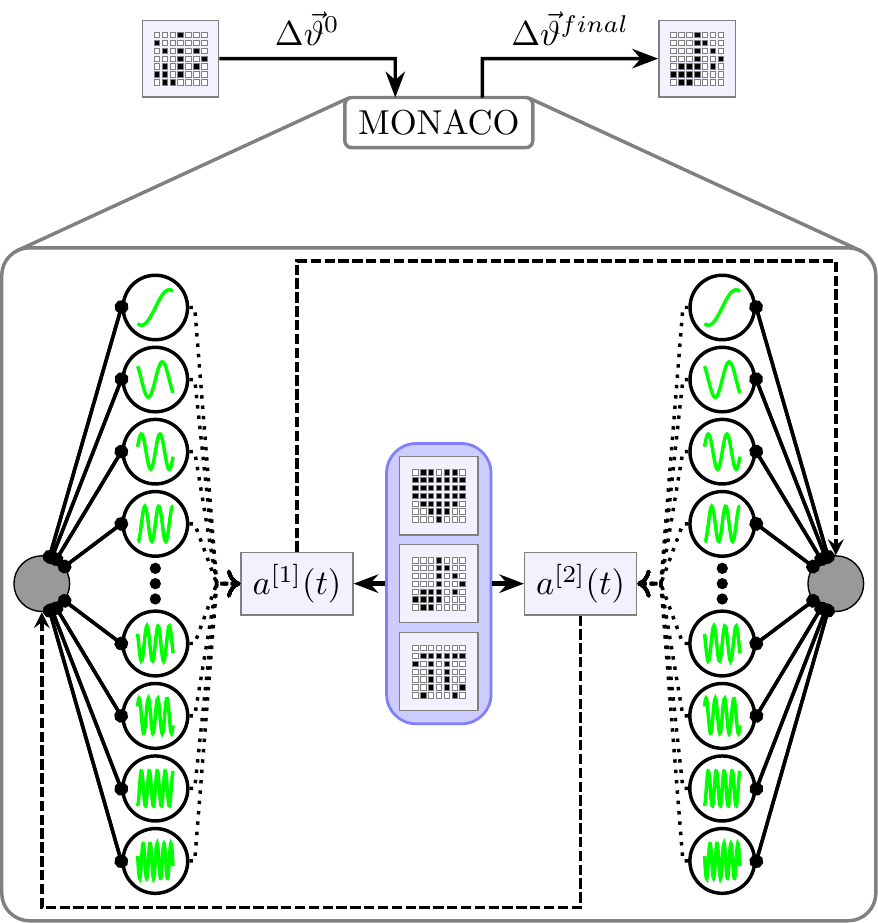}
	\caption{
	Schematics of our new MONACO-Architecture:
	Oscillators (depicted as black circles surrounding a green sine-wave) are 
	divided into two networks with the same frequency 
	distribution that are both globally coupled. 
	Coupling strength of both global couplings is modulated in time 
	 with functions $a^{[1]}(t)$ / $a^{[2]}(t)$
	 that depend on physical signals of oscillators from the other network 
	 and patterns shown in the middle.
	}
	 \label{fig:crosswise} 
\end{figure}

\newertext{The proposed} architecture consists of two identical networks of N oscillators each with equal 
frequency distribution. 
Oscillators within each of these ''subnetworks'' are globally coupled 
and the coupling strength is additionally 
modulated in time.
 For the first network,
 the {\cf{}} \footnote{Note that coupling modulations were named ''coupling functions'' by Hölzel 
 \cite{Hoelzel2011}, 
but this term is already used differently in the field.} is constructed from products of signals of the second network's oscillators 
 and vice versa.
Due to its symmetrical layout, which is visualized in Fig. \ref{fig:crosswise},
 we name the network the MONACO-Architecture: 
{\it Mirrored Oscillator Networks for Autoassociative COmputation}. 

Motivated by experiments with networks 
of electrical Van-der-Pol-oscillators \cite{Hoelzel2011,Kostorz2013}, we assume 
that the oscillators are weakly coupled in one variable, 
have sinusoidal signals and a phase response curve proportional to a cosine.
Then, the \newtext{recognition} dynamics can be reduced to a phase description
  \cite{kuramoto2012chemical}:
\begin{align}
\begin{split}
\dot{\vartheta}^{[1]}_i &=\Omega_i+\cos{\vartheta^{[1]}_i} \cdot
	a^{[2]}(t)\cdot\frac{\epsilon}{N}\sum_{j=1}^N\limits \sin{\vartheta^{[1]}_j} \\
\dot{\vartheta}^{[2]}_i &=\Omega_i+\cos{\vartheta^{[2]}_i} \cdot
	 a^{[1]}(t) \cdot\frac{\epsilon}{N}\sum_{j=1}^N\limits\sin{\vartheta^{[2]}_j} \\
{a}^{[1]}(t)&=\sum_{k,l=1}^N\limits S_{kl} \sin{\vartheta^{[1]}_k}\sin{\vartheta^{[1]}_l}\\
{a}^{[2]}(t)&=\sum_{k,l=1}^N\limits S_{kl} \sin{\vartheta^{[2]}_k}\sin{\vartheta^{[2]}_l}
\end{split}
\label{eq:phase_equations}
\end{align}
$\vartheta^{[1]}_i$ is the phase  of the i-th oscillator in the first network 
 and $\Omega_i$ its natural frequency.
In the global signal $a^{[2]}(t)\cdot\epsilon/N\cdot\sum_{j=1}^N\sin{\vartheta^{[1]}_j}$, 
 $a^{[2]}(t)$ denotes the \cf{}
generated from the second network's signals  and $\epsilon$ is a small parameter which will be shown 
to be the effective coupling strength of the averaged dynamics. 
The  amplitude perturbation is converted into a change in phase by multiplying with the phase 
response function $\cos{\vartheta^{[1]}_i}$ and the coupling matrix $\matr{\cmatrix}$ 
controls attractors of the system.

Note that the frequency distribution is the same in both networks, 
so $N$ pairs of oscillators with equal frequency exist.
For sufficiently weak coupling and specifically chosen frequencies, 
 $\cmatrix_{ij}$ 
only effectively connects oscillator pairs $i$ and $j$ and
the 
architecture can act as an autoassociative memory: 
Apart from $\Omega_{i}\neq\Omega_{j}\,\,\forall i\neq j$, 
all frequencies $\Omega_i$ must be larger than $\Omega_{max}/3$ 
and all difference frequencies $\Delta\Omega_{ij}=\Omega_i-\Omega_j$ 
must be pairwise different as shown in Appendix \ref{app:averaging}.
As we demonstrate below(Eq. \eqref{eq:theta_dynamics}), these conditions 
allow for further simplification of Eq. \ref{eq:phase_equations}.

As the oscillator pairs of equal frequency synchronize at phase differences 
$\deltheta_i=\vartheta_i^{[1]}-\vartheta_i^{[2]}$ of 
either $0$ or $\pi$ ($\pm2\pi n$) in this setup, the $\deltheta_i$ are 
easy to read out (e.g. with one signal multiplication and a low-pass filter) 
and will be our ''system state'' to be 
manipulated. The coupling matrix is chosen according to 
the Hebbian Rule \cite{hebb2005reprint}:
\begin{equation}
\cmatrix_{ij}=\sum_{m=1}^M \memvec_i^m\memvec_j^m\text{ with }\memvec_i\in\{\pm1\} \label{eq:Hebbian_rule}
\end{equation}
Then attractors will \newtext{exist for each memorized} pattern $\vec{\memvec}^m$ and its 
 inverse $-\vec{\memvec}^m$ according to the 
 following $\{ \Delta\vec{\vartheta}\mapsto\vec{\sysvec} \}$-mapping(see also Sec. \ref{sec:fixed_points}):
\begin{equation}
	\begin{matrix} 
		0 + 2\pi n \mapsto +1 \\ 
		\pi + 2\pi n \mapsto -1 
	\end{matrix}
 \label{eq:mapping_memorized_patterns}
\end{equation}
When we talk about patterns ''being attractive'', it is meant in the sense
 that attractors in $\Delta\vec{\vartheta}$ exist according to this mapping.
 
\newtext{
Assume a defective pattern $\vec{\memvec}^d$ should be recognized as 
a pattern $\vec{\memvec}^{m'}$, which is the most similar to 
$\vec{\memvec}^d$ out of $M$ correct pattern candidates $\vec{\memvec}^m$. 
For the recognition, $\vec{\memvec}^d$ is set as initial condition of the 
network according to Eq. \eqref{eq:mapping_memorized_patterns} and
the coupling matrix $\cmatrix_{ij}$ contains all correct pattern candidates as
memorized patterns according to Eq. \eqref{eq:Hebbian_rule}. As the 
defective pattern is close to the correct pattern in phase space, the 
system state will move to an attractor representing $\vec{\memvec}^{m'}$ 
and can be read out.
Note that setting initial conditions is fast and easy in the 
MONACO-architecture: As the system 
state is coded into phase differences,
 simply coupling oscillator pairs
 with negative or positive sign  according to 
 $\deltheta_i=-\memvec_i^d\cdot E\sin{\deltheta_i}$ and $E\gg\epsilon$ 
 for a short time $T_{init}\ll 1/\epsilon$ 
 ensures a correct initialization. 
}

\newtext{
If instead of an erroneous pattern only a small correct part of a pattern 
is known, missing pixel in $\vec{\memvec}^d$ can be filled with $+1$ or 
$-1$ with equal probability. Afterwards, recognition 
 is performed as above.
}

\begin{figure}[h!]
\centering
\includegraphics{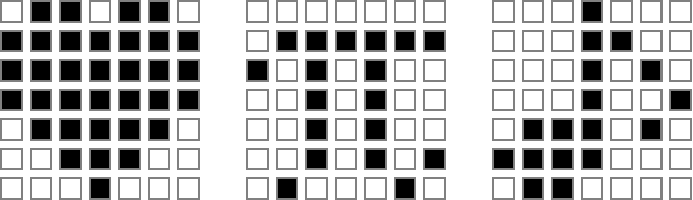}
\caption{
Non-orthogonal patterns with 49 pixels that were used as memorized patterns $\vec{\memvec}^m$ in 
simulations for Fig. \ref{fig:correct_matching}, Fig. \ref{fig:failing_simulation} 
and the statistics in Sec. \ref{sec:boa_a_matching_crit}.
$\memvec_i^m=+1$ is visualized as a black pixel and white pixels correspond 
 to $\memvec_i^m=-1$. 
}
\label{fig:output_patterns}
\end{figure}

\begin{figure}
\centering
\includegraphics{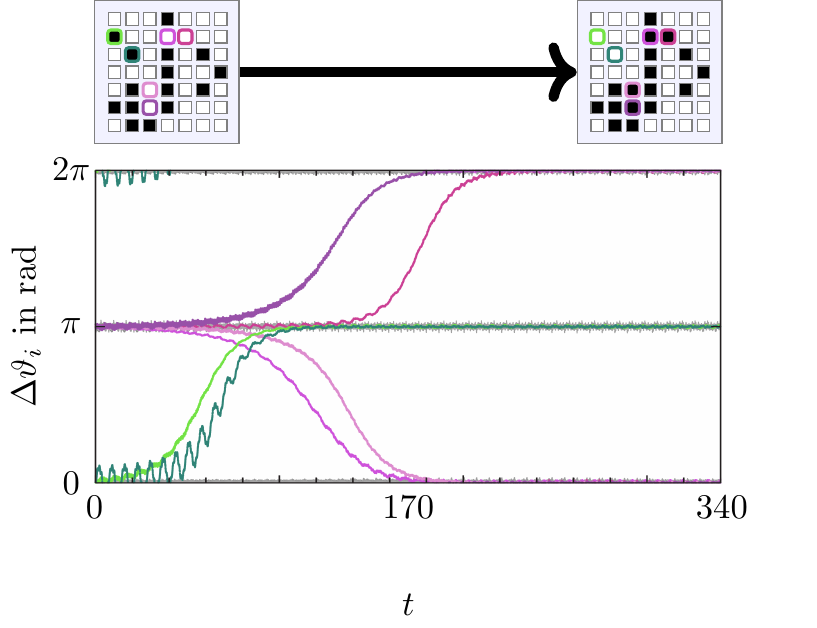}
\caption{Successful \newtext{Recognition: A} binary pattern with 6 erroneous pixels (colored and thick) shown on the top left 
	is correctly recognized as one of 3 memorized patterns shown in Fig. 
	\ref{fig:output_patterns}. 
	White pixels 
	are mapped onto $\deltheta_i = \pi$
	 and black pixels 
	 correspond to $\deltheta_i = 0\text{ or }2\pi$.
	 Green trajectories correspond to pixels that are erroneously black,  
	 so they should change from $\deltheta_i =0$ 
	 (or $2\pi$) to \newtext{$\deltheta_i =\pi$,}
	  which they do successfully. Similarly, pink trajectories 
	  correctly change from $\pi$ 
	  to $0$ (or $2\pi$) while gray trajectories, 
	   corresponding to already correct pixels, do not change.
	   For simulation details,
	 see Sec. \ref{sec:sim_details}. 
	}
	\label{fig:correct_matching}
	\label{fig:successful_simulation}
\end{figure}

Phase differences $\deltheta_j$ from an exemplary simulation 
of the phase dynamics (Eq. \eqref{eq:phase_equations}) are shown in 
Fig. \ref{fig:correct_matching}
for $N=49$ oscillator pairs \newtext{and 6 defective pixels}. The memorized 
patterns $\vec{\memvec}^m$ used 
 are visualized in Fig. \ref{fig:output_patterns}
  and are not orthogonal in the sense that 
  $\langle \vec{\memvec}^{m_1},\vec{\memvec}^{m_2}\rangle\neq 0$ $\forall m_1,m_2$ and $m_1\neq m_2$ 
  ($\langle,\rangle$ denotes the standard scalar \newtext{product.).}
\newtext{The }
erroneous phase differences change to represent the correct \textmusicalnote-shaped output 
\newtext{pattern.} 

\begin{figure}
\centering
\includegraphics{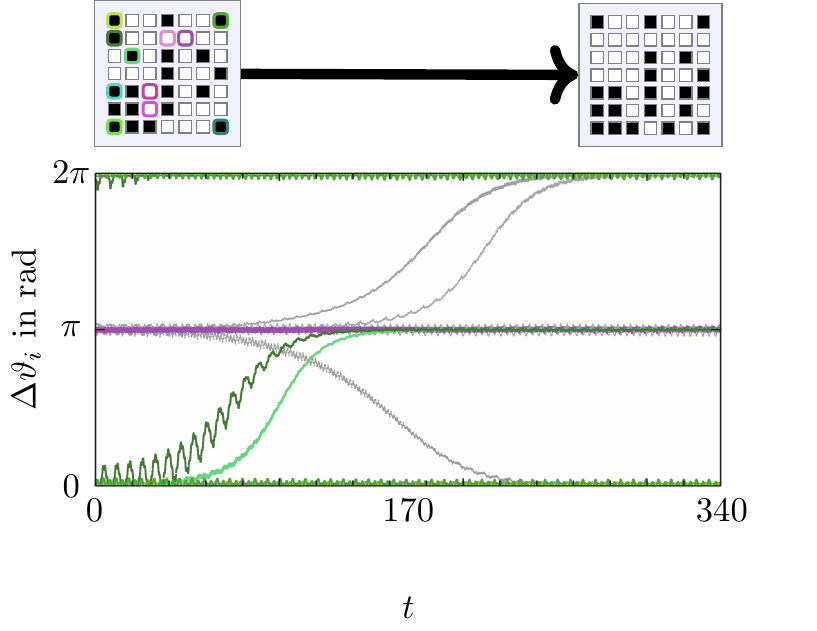}
	\caption{\newtext{Recognition fails} due to too many defects:
	A binary pattern with 11 erroneous pixels (colored) shown on the top left 
	should be recognized as one of 3 memorized patterns shown in Fig. 
	\ref{fig:output_patterns}. 
	However, the recognition fails:
	  Pink trajectories corresponding to pixels that are erroneously white  
	  should change from $\deltheta_i=\pi$ to  $\deltheta_i=0$ (or $2\pi$)
	   during the recognition process, 
	  but do not. 
	  Likewise, only 2 green trajectories change from 
	  $0$ (or $2\pi$) to $\deltheta_i=\pi$, although all 7 
	  represent black pixels that should change to white. Additionally 3 gray trajectories,
	   whose corresponding pixels are already correct, change to wrong values.
	   The system settles at the pattern shown on the top right, 
	   which is none of the memorized patterns.
	    For simulation details,
	 see Sec. \ref{sec:sim_details}. 
	}
	\label{fig:failing_simulation}
\end{figure}

However, the recognition process can fail if the number of erroneous pixels is too large. 
A failed \newtext{recognition is} shown in Fig. \ref{fig:failing_simulation}: The system state
moves to an unknown attractor which corresponds to none of the 
$\vec{\memvec}^m$. In order to predict recognition success, a simple 
criterion is derived and tested in Sec. \ref{sec:boa_a_matching_crit}. 

\newertext{
Before analyzing the dynamics, we want to point out that
 the coupling matrix $\matr{S}$ does not need to be wired explicitly, which would require 
$\mathcal{O}(N^2)$ connections. By rewriting both {\cf{}}s as squares of scalar products instead,
they can be generated with $\mathcal{O}(N\cdot M)$ 
connections only: }
\newtext{
\begin{align}
	a^{[1/2]}(t)&=\sum_{k,l=1}^N\limits 
	\sum_{m=1}^M\limits \memvec_k^m\memvec_l^m \sin{\vartheta^{[1/2]}_k}\sin{\vartheta^{[1/2]}_l} \nonumber\\
	&=\sum_{m=1}^M\limits \Big ( \sum_{j=1}^N\limits \memvec_j^m \sin{\vartheta^{[1/2]}_j} \Big)^2 
	\label{eq:simplified_coupling_modulations}
\end{align}
}
\newertext{
Whenever MONACO is used as an autoassociative memory as presented here, 
$a^{[1]}(t)$ and $a^{[2]}(t)$ should therefore always be constructed according to
Eq. \eqref{eq:simplified_coupling_modulations} instead of Eq. \eqref{eq:phase_equations}.
}
Depending on usage, the $\vec{\memvec}^m$ can be hardwired or changed for each recognition process.

\section{Analysis of the dynamics}
\label{sec:fixed_points}
\subsection{Simplification of the evolution equations}
Prior to determining attractors, we simplify the phase equations (Eq. \eqref{eq:phase_equations}) 
with the technique of averaging \cite{verhulst1996nonlinear}:

 The right hand sides of Eq. \eqref{eq:phase_equations}
consist of many different frequency components.
 If the coupling strength $\epsilon$ is sufficiently small,
  larger frequencies average out on times much smaller than the 
 largest timescale and the smallest frequencies dominate the dynamics:
 \begin{align*}
	\dot{\vartheta}^{[1]}_i 
		&\approx\Omega_i
		+ \frac{\epsilon M}{8 N}\sinAv{2\deltheta_i} \\
		&-\frac{\epsilon}{4N}\sum_{j=1}^N\limits\cmatrix_{ij}   
			\left [ \sinAv{\deltheta_i+\deltheta_j} + \sinAv{\deltheta_i-\deltheta_j} \right ] \\
	\dot{\vartheta}^{[2]}_i 
		&\approx\Omega_i
		- \frac{\epsilon M}{8N}\sinAv{2\deltheta_i} \\
		&+\frac{\epsilon}{4N}\sum_{j=1}^N\limits\cmatrix_{ij} 
		\left [ \sinAv{\deltheta_i+\deltheta_j} + \sinAv{\deltheta_i-\deltheta_j} \right ]
\end{align*}
The lengthy averaging calculation is shown in Appendix \ref{app:averaging} and includes 
 restrictions on the frequency distribution of the oscillators.
 Using a trigonometric theorem, we can express our equation system with the phase differences 
 $\deltheta_i$ only:
 \begin{align}
 \Delta\dot{\vartheta}_i&=\dot{\vartheta}^{[1]}_i -\dot{\vartheta}^{[2]}_i \approx \nonumber \\
		&-\frac{\epsilon}{2N}\sum_{j=1}^N\limits\cmatrix_{ij}   
			\left [ \sinAv{\deltheta_i+\deltheta_j} + \sinAv{\deltheta_i-\deltheta_j} \right ]  \nonumber \\
	&+ \frac{\epsilon M}{4 N}\sinAv{2\deltheta_i} \nonumber \\
	\Delta\dot{\vartheta}_i	&=-\frac{\epsilon}{N} \sin{\deltheta_i}
	\Big (\sum_{j=1}^N\limits\cmatrix_{ij}\cos{\deltheta_j} - \frac{M}{2}\cos{\deltheta_i}\Big )
	\label{eq:theta_dynamics}
\end{align}
This is the main evolution equation that governs the dynamics of the architecture.

\subsection{Fixed points and their stability}  
At fixed points $\Delta\vec{\vartheta}^*$ of the dynamics, 
all velocity components $\Delta\dot{\vartheta}_i$ must vanish. 
Depending on which factor in Eq. \eqref{eq:theta_dynamics} vanishes,
pixel indices can be sorted into \newertext{two} sets $\isolatedSet$ and $\unstableSet$:\\
\begin{align}
	&\text{\textbullet}\,\, i \in \isolatedSet \IF \sin{\deltheta_i^*} =  0
	\IF  \deltheta_i^* \in \{0,\pi\}+2\pi n \nonumber \\
	&\text{\textbullet}\,\, i \in \unstableSet 
	\IF \sum_{j=1}^N\limits\cmatrix_{ij}\cos{\deltheta_j^*}-\frac{M}{2}\cos{\deltheta_i^*}= 0
		\label{eq:def_unstableSet}
\end{align}
We show in Appendix \ref{app:potential} that all fixed points with indices in $\unstableSet$ 
are unstable. Therefore, all attractors 
are well-separated fixed points with $i \in \isolatedSet\,\,\forall i$.\\

Only fixed points with 	$\sin{\deltheta_i^*} =  0$ $\forall i$
 and $\sum_{j=1}^N\cmatrix_{ij}\cos{\deltheta_j^*}-M/2\cos{\deltheta_i^*}\neq 0$ remain as candidates 
 for attractors.

The stability of fixed points can generally be examined by linearizing 
the dynamics around the fixed point by 
evaluating the eigenvalues of  
the Jacobian $J_{ik}=\partial\Delta\dot{\vartheta}_i / \partial\phasediff_k$ at the fixed point $\Delta\vec{\vartheta}^*$:
\begin{align*} J_{ik}
	=&-\frac{\epsilon}{N}\cos{\phasediff_i}\delta_{ik}\Bigg ( \sum_{j=1}^N\limits\cmatrix_{ij}\cos{\deltheta_j} - \frac{M}{2}\cos{\deltheta_i}\Bigg ) \\
	&- \frac{\epsilon}{N}\sin{\phasediff_i}\Bigg ( -\cmatrix_{ik}\sin{\phasediff_{\newtext{k}}} + \frac{M}{2}\delta_{ik}\sin{\deltheta_i}\Bigg )
\end{align*}
As $i \in \isolatedSet\,\forall i$ implies $\sin{\deltheta_i^*} =  0\,\forall i$, the second term vanishes:
 \[J_{ik}(\Delta\vec{\vartheta}^*)
=-\delta_{ik}\frac{\epsilon}{N}\cos{\phasediff_i^*}\Bigg ( \sum_{j=1}^N\limits\cmatrix_{ij}\cos{\deltheta_j^*} - \frac{M}{2}\cos{\deltheta_i^*}\Bigg )
\]
$\matr{J}$ is a diagonal matrix, therefore eigenvectors $\hat{e}_i$ are the standard base 
 with the following eigenvalues:
\[\lambda_i=-\frac{\epsilon}{N}\cos{\phasediff_i^*}\Bigg ( \sum_{j=1}^N\limits\cmatrix_{ij}\cos{\deltheta_j^*} - \frac{M}{2}\cos{\deltheta_i^*}\Bigg ) \] 

We can simplify the analysis further by defining ''pattern coordinates'' $\vec{\sysvec}$ 
with $\sysvec_i=\cos{\deltheta_i}$ as generalization of Eq. \eqref{eq:mapping_memorized_patterns} 
 and inserting the definition of the coupling matrix $\matr{\cmatrix}$: 
\begin{align}
\lambda_i &= -\frac{\epsilon}{N}\Bigg ( \sum_{m=1}^M\limits  \memvec_i^m \sysvec^{*}_i 
\sum_{j=1}^N  \memvec_j^m \sysvec^{*}_j - \frac{M}{2}{\sysvec^{*}_i}^2 \Bigg ) \nonumber\\
\lambda_i &= -\frac{\epsilon}{N}\Bigg ( \sum_{m=1}^M\limits  \memvec_i^m \sysvec^{*}_i \left < \vec{\memvec}^m,\vec{\sysvec}^{*}\right > - \frac{M}{2}\Bigg )
\label{eq:eigenvalues}
\end{align}
The signs of the eigenvalues determine the stability: 
Positive eigenvalues denote growing perturbations along the 
corresponding eigendirection, while negative eigenvalues indicate decay.
Therefore, all fixed points with $\lambda_i<0\,\forall i$ are isolated attractors:
\begin{align}
	\sum_{m=1}^M\limits  \memvec_i^m \sysvec^{*}_i \left < \vec{\memvec}^m,\vec{\sysvec}^{*}\right > > \frac{M}{2} 
	&& \wedge && \sysvec^{*}_i \in \{\pm 1\} \label{eq:attractors}
\end{align}

\textit{ Memorized patterns map to isolated attractors},
 if inter-pattern scalar products are sufficiently small.
If patterns $\vec{\memvec}^m$ are orthogonal, inter-pattern scalar products vanish completely:
 \begin{align}
 \lambda_i(\vec{\memvec}^{m'}) 
 &= -\frac{\epsilon}{N}\left (\sum_{m=1}^M\limits  \memvec_i^m \memvec_i^{m'} 
 \left < \vec{\memvec}^m,\vec{\memvec}^{m'}\right > - \frac{M}{2}\right ) \nonumber\\
  &= -\frac{\epsilon}{N}\left (\sum_{m=1}^M\limits  \memvec_i^m \memvec_i^{m'} 
 \delta_{mm'}N- \frac{M}{2}\right ) \nonumber \\
 &=-\epsilon\left (1-\frac{M}{2 N}\right) <0\quad\forall i \nonumber\\
\Leftrightarrow \quad M&<2N \label{eq:orthogonal_stability_criterion}
\end{align}
Not more than $N$ orthogonal patterns can exist
($\vspan(\vec{\memvec}^m)\leq N$, but $\vspan(\vec{\memvec}^m) = M$ for linear independent patterns.), 
so $M<2N$ is always fulfilled and 
orthogonal patterns are guaranteed to be stable.

For general $\vec{\memvec}^m$, we get 
\begin{align*}
\lambda_i(\vec{\memvec}^{m'}) &\overset{!}{<}0 \\
-\epsilon  -\frac{\epsilon}{N}\Bigg (\sum_{m\neq m'}^M\limits  \memvec_i^m \memvec_i^{m'} 
 \left < \vec{\memvec}^m,\vec{\memvec}^{m'}\right > - \frac{M}{2}\Bigg )
&<0 \\
-\sum_{m\neq m'}^M\limits  \memvec_i^m \memvec_i^{m'} 
 \left < \vec{\memvec}^m,\vec{\memvec}^{m'}\right > &< N - \frac{M}{2}\text{.}
\end{align*}
As we want a criterion to ensure that all memorized patterns are attractors, 
we must exclude that any eigendirection of any pattern becomes unstable:
\begin{align}
 \max_{i,m'}\limits \Bigg ( -\sum_{m\neq m'}^M\limits  \memvec_i^m \memvec_i^{m'} 
 \left < \vec{\memvec}^m,\vec{\memvec}^{m'}\right > \Bigg ) &< N - \frac{M}{2} \nonumber \\
\Sigma_{max} \overset{!}{=} \max_{m'}\limits \Bigg ( \sum_{m\neq m'}^M\limits  
 \left | \left < \vec{\memvec}^m,\vec{\memvec}^{m'}\right > \right | \Bigg ) &< N - \frac{M}{2} 
 \label{eq:general_stability_criterion}
\end{align}

Additionally, if the $\vec{\memvec}^m$ are attractors, 
their \textit{inverses will be attractors as well}
 because their eigenvalues are identical:
\begin{align*}
	\lambda_i(-\vec{\memvec}^{m'}) &= 
-\frac{\epsilon}{N}\Bigg [ \sum_{m=1}^M\limits  \memvec_i^m (-\memvec_i^{m'}) 
\left < \vec{\memvec}^m,(-\vec{\memvec}^{m'})\right > - \frac{M}{2}\Bigg ]\\
 &= 
-\frac{\epsilon}{N}\Bigg ( \sum_{m=1}^M\limits  \memvec_i^m \memvec_i^{m'} 
\left < \vec{\memvec}^m,\vec{\memvec}^{m'}\right > - \frac{M}{2}\Bigg )\\
&=\lambda_i(\vec{\memvec}^{m'})
\end{align*}

Moreover, there are further spurious attractors that do not represent one of the 
$\vec{\sysvec}^m$, but they are difficult 
to describe. If the initial pattern does not start in the basin of attraction 
of an $\vec{\memvec}^m$, the output of the system will be one of these attractors. 
Therefore, stability is not sufficient for recognition success and we 
have to derive a criterion from the basins of attraction. 
However, first we derive a more common criterium for the network
 capacity that can be compared in different network architectures. 

\newtext{
\subsection{Error-free capacity}
\label{subsec:errorfree}
The \emph{error-free capacity} $M_{max}(N)/N$ is a measure 
 for the amount of memorized patterns $\vec{\memvec}^{m}$ that 
 can be stored in a  given network 
  while \newertext{any pattern} can still be retrieved without errors.
\newertext{Specifically, we determine the maximum number of patterns
 $M_{max}(N)$ so $P(\vec{\memvec}^{m'}\text{ is stable})\to1$ for $M<M_{max}(N)$
 and $P(\vec{\memvec}^{m'}\text{ is stable})\to 0$ for $M>M_{max}(N)$.}
\newertext{Similar to approaches for other architectures, 
we derive $M_{max}(N)$ }
in a probabilistic manner for random memorized patterns 
 with $P(\memvec_i^m=+1)=P(\memvec_i^m=-1)=0.5\,\forall m,i$ in the limes 
 $N\to\infty$.\\
\newertext{First, we} simplify the
rescaled Jacobian $\matr{\tilde{J}}= \matr{J}/\epsilon$ at a memorized pattern $\vec{\memvec}^{m'}$: 
\begin{align*}
\tilde{J}_{ik}(\vec{\memvec}^{m'}) 
&= -\delta_{ik}\frac{1}{N}\Bigg ( \sum_{m=1}^M\limits  \memvec_i^m \memvec^{m'}_i \sum_{j=1}^N  \memvec_j^m \memvec^{m'}_j - \frac{M}{2}{\memvec^{m'}_i}^2 \Bigg )\\
&= -\delta_{ik}\frac{1}{N}\Bigg ( \sum_{
		\begin{subarray}{l}
			m=1\\
			m\neq m'
		\end{subarray}}^M\limits  
		\memvec_i^m \memvec^{m'}_i \sum_{j=1}^N  \memvec_j^m \memvec^{m'}_j + N - \frac{M}{2}\Bigg ) \\
&\begin{aligned}
	= -\delta_{ik}\frac{1}{N}\Bigg ( \sum_{
		\begin{subarray}{l}
			m=1\\
			m\neq m'
		\end{subarray}}^M\limits 
		\sum_{
		\begin{subarray}{l}
			j=1\\
			j\neq i
		\end{subarray}}^N 
		\memvec_i^m \memvec^{m'}_i   \memvec_j^m \memvec^{m'}_j \\
		 + (M-1)
		  + N - \frac{M}{2}\Bigg ) 
	\end{aligned}\\
\matr{\tilde{J}}(\vec{\memvec}^{m'})&=-\left(1 + \frac{M-2}{2N} \right )\matr{I} + \matr{D}
\end{align*}
Here, $\matr{I}$ is the identity matrix and 
\begin{equation*}
D_{ik}=-\delta_{ik}\frac{1}{N}\sum_{
		\begin{subarray}{l}
			m=1\\
			m\neq m'
		\end{subarray}}^M 
		\sum_{
		\begin{subarray}{l}
			j=1\\
			j\neq i
		\end{subarray}}^N 
		\memvec_i^m \memvec^{m'}_i   \memvec_j^m \memvec^{m'}_j\text{.}
\end{equation*}
 As $\matr{\tilde{J}}$, $\matr{I}$ and $\matr{D}$ are diagonal, 
 \begin{align*}
 \lambda_{max}(\matr{\tilde{J}})
 &=-\left(1 + \frac{M-2}{2N} \right )+\lambda_{max}(\matr{D}) \\
 &=-\left(1 + \frac{M-2}{2N} \right )+\max_i D_{ii}
 \text{.}
 \end{align*}
Then, the stability condition can be expressed as function of $\max_i D_{ii}$ 
alone:
\begin{align}
	&&\lambda_{max}(\matr{J})&<0 \nonumber \\
	\newertext{\Leftrightarrow} &&\lambda_{max}(\matr{\tilde{J}}) &<0 \nonumber \\
	\newertext{\Leftrightarrow} && \max_i D_{ii}&<1 + \frac{M-2}{2N} \label{eq:stability_cond}
\end{align}
 The following lemma concerning this largest eigenvalue  
 $\lambda_{max}(\matr{D})=\max_i D_{ii}$ has been proven in
\cite{nishikawa2004_second_order} as Lemma 6 under the assumption that 
all pixels of all memorized patterns $\vec{\memvec}^m$ are randomly chosen 
with probability $P(+1)=P(-1)=0.5$: \newertext{(All occurring logarithms are natural.)}\\
\begin{lemma}
$\,$\\Let $x>0$, and
\begin{equation*}
\bar{\limSymbol}=\limsup_{N\to\infty}\limits \frac{M(N)\log{(N)}}{N}\text{,}
\quad\,\,\ubar{\limSymbol}=\liminf_{N\to\infty}\limits \frac{M(N)\log{(N)}}{N}\text{.}
\end{equation*}
 If $\bar{\limSymbol}<x^2/2$, then 
$P(max_i D_{ii}\geq x)\to 0$ as $N \to \infty$.\\
If $\ubar{\limSymbol}>x^2/2$, 
then $P(max_i D_{ii}\geq x)\to 1$ as $N\to\infty$.
\end{lemma}
According to Eq. \eqref{eq:stability_cond}, 
$\vec{\memvec}^{m'}$ is stable for $N\to\infty$ and $M>1$ if $\max_i D_{ii}<1$  
is fulfilled. Therefore, we are interested in the probability  
$P(max_i D_{ii}\geq 1)$ and choose $x=1$. \\
$P(\vec{\memvec}^{m'}\text{ stable})=1-P(max_i D_{ii}\geq 1)\to 1$ 
 if $\bar{\limSymbol}<1/2$ and $P(\vec{\memvec}^{m'}\text{ unstable})=P(max_i D_{ii}\geq 1)\to 1$ 
 if $\ubar{\limSymbol}>1/2$, which implies 
 \begin{align}
	M_{max}(N)&=\frac{N}{2\log{(N)}} \nonumber \\
	\Leftrightarrow \frac{M_{max}(N)}{N}&=\frac{1}{2\log{(N)}} \label{eq:errorfree_capacity}
 \end{align}
Note that other capacity measures exist, such as the 
\emph{loading rate}, which describes the fraction $M_{max}/N$ under the 
assumption that attractors for each pattern do exist, but might be shifted, 
so retrieved patterns might have some errors. Therefore, 
the \emph{error-free capacity} always is a lower bound in the loading rate. 
While these probabilistic measures are useful for comparing architectures, 
their validity is constrained in reality: 
 Real networks are of finite size and memorized 
 patterns need not be chosen randomly. 
Bounds on guaranteed stability were derived \newertext{in Eq. \eqref{eq:general_stability_criterion}}
 and a criterion for guaranteed recognition is derived in Section \ref{sec:boa_a_matching_crit}.
} 
\begin{figure}
	\includegraphics[width=8.5cm]{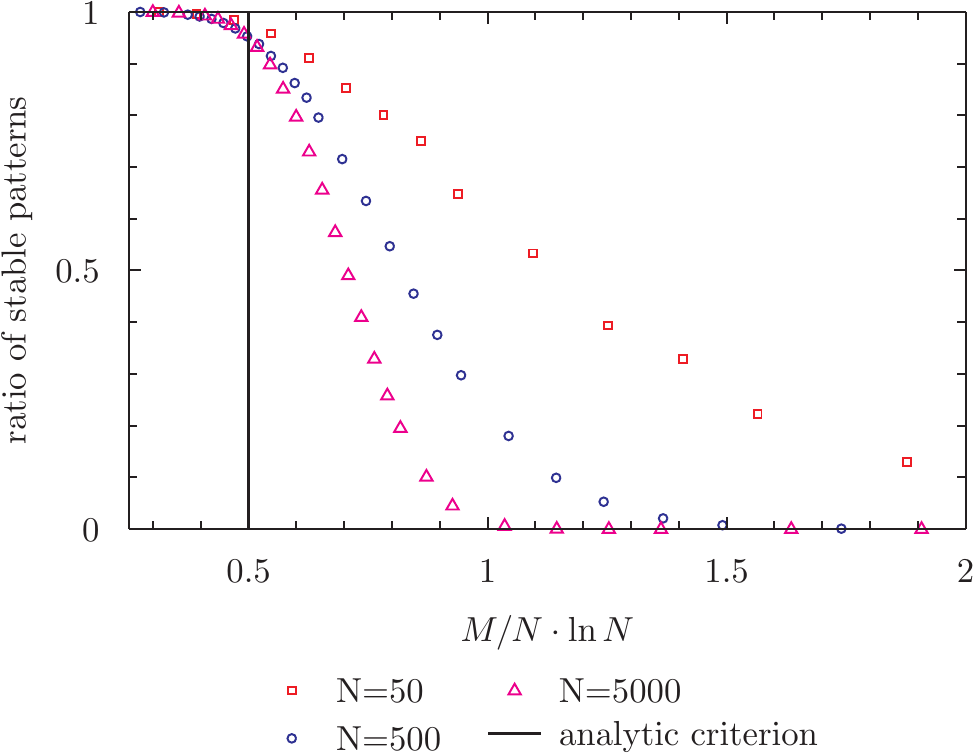}
	\caption{The analytic criterion for the error-free capacity Eq. 
	\eqref{eq:errorfree_capacity}  
	is compared to 
	explicit evaluation of the patterns' eigenvalues. 
	For each datapoint, 1000 sets of pattern were created randomly 
	with $P(\memvec_i^m=+ 1)=P(\memvec_i^m=- 1)=0.5$ and the stability
	of each pattern 
	was determined with Eq. \eqref{eq:attractors}.}
\end{figure}

\subsection{Intuitive explanation of the recognition mechanism}
The results of the fixed point analysis allow a more intuitive view of Eq. \eqref{eq:theta_dynamics} 
by partially expressing the system state in pattern coordinates $\vec{\sysvec}$ 
with $\sysvec_i=\cos{\deltheta_i}$:
\begin{align*}
	\Delta\dot{\vartheta}_i	&=-\frac{\epsilon}{N} \sin{\deltheta_i}
	\Bigg( \sum_{j=1}^N\limits\cmatrix_{ij}\cos{\deltheta_j} - \frac{M}{2}\cos{\deltheta_i}\Bigg ) \\
	&=-\frac{\epsilon}{N} \sin{\deltheta_i}
	\Bigg ( \sum_{j=1}^N\limits \sum_{m=1}^M\limits \memvec_i^m \memvec_j^m \sysvec_j - \frac{M}{2}\sysvec_i \Bigg ) \\
	&=  -\sin{\deltheta_i}  \cdot \frac{\epsilon}{N} 
	\Bigg (  \sum_{m=1}^M \limits \memvec_i^m \left < \vec{\memvec}^m,\vec{\sysvec} \right >
	 - \frac{M}{2}\sysvec_i \Bigg ) 
\end{align*}
\newertext{Let $\vec{\memvec}^{m'}$ be the memorized pattern the system state $\vec{\sysvec}$ is closest to:
\begin{align*}
	\Delta\dot{\vartheta}_i	&=-\sin{\deltheta_i}  \cdot \memvec_i^{m'} \cdot \frac{\epsilon}{N} \cdot \\
	\cdot&\Bigg ( \left < \vec{\memvec}^{m'},\vec{\sysvec} \right > 
	+ \sum_{
\begin{subarray}{l}
m=1\\
m\neq m'
\end{subarray}}^M
 \limits \memvec_i^{m'}\memvec_i^m \left < \vec{\memvec}^m,\vec{\sysvec} \right >
	 - \frac{M}{2} \memvec_i^{m'} \sysvec_i \Bigg )
\end{align*}
Now assume the system state $\vec{\sysvec}$ is sufficiently close to 
 $\vec{\memvec}^{m'}$:
Then  $ \langle\vec{\memvec}^{m'},\vec{\sysvec}\rangle$ is 
larger than the sum of all other terms in parentheses. Hence,}  
the fixed points and their stability are the same as 
in $\text{d}/\text{d}t \,\deltheta_i= -\memvec_i^{m'}\sin{\deltheta_i} $.
 If $\memvec_i^{m'}=+1$, 
 $\deltheta^*_i=0$ is stable and $\deltheta^*_i=\pi$ is unstable and 
 vice-versa for $\memvec_i^{m'}=-1$, so $\lim_{t\rightarrow\infty}\sysvec_i=\cos{\deltheta_i^*}=\memvec_i^{m'}$. 
 
 From another point of view, the system ''defines'' ''relative closeness'' to 
 memorized patterns by comparing their projections onto the system state 
 $\vec{\sysvec}$. This fails, however, if the scalar products are of 
 comparable size: Then the distribution of the $\memvec_i^{m}$ matters 
 for each pixel, which leads to spurious attractors 
unequal to all $\vec{\memvec}^m$. 
\newertext{Note that }the $- M\sysvec_i/2$-term does not really 
contribute to the recognition mechanism. While it increases eigenvalues 
of all stable fixed points slightly, therefore reducing stability \newertext{(see Eq. \eqref{eq:eigenvalues})},
 it does not influence the basins of attraction much, as we will 
 illustrate in the next section.
\section{Basins of attraction and guaranteed recognition}
\label{sec:boa_a_matching_crit}
We have a firm understanding of the system now and can guarantee 
that the chosen patterns $\vec{\memvec}^m$ are attractive. However, 
 we cannot guarantee 
recognition success yet: The system state might relax to the additional
 unwanted attractors described by Eq. \eqref{eq:attractors} or even worse, the basins of 
 attraction of the $\vec{\memvec}^m$ might be malformed, 
 leading to a $\vec{\memvec}^m$ whose projection on the defective 
 pattern is not the largest.
\subsection{Lower bound on the basins of attraction}
Matching success is guaranteed if the defective starting pattern is in
 the basin of attraction of the correct memorized pattern $\vec{\memvec}^{m'}$. 
 A lower bound on the basin of attraction can be derived by proofing the
 following lemmata:
 \begin{enumerate}
	\item Surfaces of constant projection 
	 on the correct memorized pattern $\vec{\memvec}^{m'}$
	confine the system state to larger projections
	 if the initial projection is sufficiently large.  
	\item $\vec{\memvec}^{m'}$ is the only attractor 
	inside this confined space.
 \end{enumerate}
As the system state cannot leave the confined space, it has to settle
on $\vec{\memvec}^{m'}$ as the only attractor. Therefore, the confined 
space is part of $\vec{\memvec}^{m'}$s basin of attraction.

\subsubsection{Transformation to $\vec{\sysvec}$-space}
For our following discussion, we will transfer the  
$\Delta\vec{\vartheta}_i$-dynamics (Eq. \eqref{eq:theta_dynamics}) completely 
into the ''\patterncoordinates'' $\vec{\sysvec}$ with $\sysvec_i = \cos\deltheta_i$, 
which are a generalization 
of the mapping of the memorized patterns 
$\vec{\memvec}^m$.
\begin{align}
	\dot{\sysvec_i}&=\dot{(\cos{\phasediff_i})} \nonumber \\
		&=\frac{\partial\cos{\phasediff_i}}{\partial\phasediff_i}\frac{\partial\phasediff_i}{\partial t} \nonumber  \\
		&
		\begin{aligned}
			=-\sin{\phasediff_i}\Bigg[ -\frac{\epsilon}{N}
			\sin{\phasediff_i}\Bigg ( \sum_{j=1}^N\limits\cmatrix_{ij}\cos{\deltheta_j} \\
			- \frac{M}{2}\cos{\deltheta_i} \Bigg ) \Bigg ] 
		\end{aligned} \nonumber  \\
		&
		\begin{aligned}
		=\frac{\epsilon}{N}\sin^2{\phasediff_i} \Bigg ( \sum_{j=1}^N\limits\sum_{m=1}^M \memvec_i^m \memvec_j^m\cos{\deltheta_j}\\
		 - \frac{M}{2}\cos{\deltheta_i}\Bigg ) 
		\end{aligned}  \nonumber \\
		&
		\begin{aligned}
		=\frac{\epsilon}{N}(1-\cos^2{\phasediff_i})\Bigg ( \sum_{j=1}^N\limits\sum_{m=1}^M \memvec_i^m \memvec_j^m\cos{\deltheta_j} \\
		- \frac{M}{2}\cos{\deltheta_i}\Bigg )
		\end{aligned}   \nonumber \\
		&=\frac{\epsilon}{N}(1-\sysvec_i^2)\Bigg ( \sum_{m=1}^M   \memvec_i^m \sum_{j=1}^N \memvec_j^m \sysvec_j  - \frac{M}{2}\sysvec_i \Bigg )  \nonumber \\
	\dot{\sysvec_i}&=\frac{\epsilon}{N}(1-\sysvec_i^2){}
	\Bigg ( \sum_{m=1}^M   \memvec_i^m 
	\left <  \vec{\memvec}^m,\vec{\sysvec} \right >
	 - \frac{M}{2}\sysvec_i \Bigg ) \label{eq:alpha_dynamics}
\end{align}
Note that although the mapping between $\deltheta_i$ and $\sysvec_i$ 
is \emph{not} injective, the transformation is still valid: 
Eq. \eqref{eq:theta_dynamics} 
is mirror-symmetric to $0+\pi n\text{ with }n\in\mathbb{N}$,
 so space can be divided into regions separated by $\deltheta_i=[0,\pi]+2\pi n$ 
or $\deltheta_i=[\pi,2\pi]+2\pi n$ 
in every $i$ and flow lines in each region are mapped onto the 
same $\vec{\sysvec}$-coordinates. As 
the flow across the boundaries of these hypercubes is zero,
it is not necessary to consider the periodicity of the flow.
From another point of view, the ambiguity of attractors in 
$\Delta\vec{\vartheta}$ is removed in the $\vec{\sysvec}$-coordinates. 
As the dynamics of $\vec{\sysvec}$ do not depend on the sign or periodicity of 
$\Delta\vec{\vartheta}$, it is a more natural coordinate 
for the autoassociative memory.
\subsubsection{Confinement by hypersurfaces of constant projection}
Let's consider a hypersurface of constant projection on the correct output 
pattern $\vec{\sysvec}^{m'}$:
In the \patterncoordinates the equation
$\langle \vec{\sysvec}, \vec{\memvec}^{m'} \rangle = C$ 
describes a hyperplane that divides 
the N-dimensional hypercube of all possible patterns 
into patterns with a projection larger or smaller than $C$. 
If  
projections on $\vec{\sysvec}^{m'}$ do not decrease for all
points on the surface, the system 
state can only move tangential to the hyperplane or towards larger projections.
(Movement tangential to the hyperplane is 
in fact impossible with a slightly stricter condition, as shown further below.)
\begin{align}
\frac{\text{d}}{\text{d}t}\frac{1}{\left | \vec{\memvec}^{m'} \right |}
	\big < \vec{\sysvec},\vec{\memvec}^{m'}\big > &\geq 0 \nonumber \\
\frac{\text{d}}{\text{d}t} \big < \vec{\sysvec},\vec{\memvec}^{m'}\big > 
	 =\big < \dot{\vec{\sysvec}},\vec{\memvec}^{m'}\big > &\geq 0 \nonumber \\
\sum\limits_{i=1}^N (1-\sysvec_i^2)\memvec_i^{m'}\Bigg ( \sum_{m=1}^M   \memvec_i^m \left <  \vec{\memvec}^m,\vec{\sysvec} \right > 
- \frac{M}{2}\sysvec_i \Bigg ) &\geq 0 \label{eq:crit_proj>0_2}
\end{align}
If  Eq. \eqref{eq:crit_proj>0_2} is fulfilled for all $\vec{\sysvec}$ 
on a hypersurface 
$\langle \vec{\sysvec}, \vec{\memvec}^{m'} \rangle = C$, it confines
 the system state.
However, to exclude additional attractors  besides $\vec{\memvec}^{m'}$ 
in the confined space is 
difficult with Eq. \eqref{eq:crit_proj>0_2} and a good criterion for 
guaranteed recognition
 should neither  depend  on the hyperplanes nor on  
 the specific pixels of $\vec{\sysvec}$ or the memorized patterns 
 $\vec{\memvec}^m$.
 Therefore, we employ a series of {\it worst-case approximations 
 and upper bounds}:

Eq. \eqref{eq:crit_proj>0_2} is fulfilled if all single summands are greater than zero.
Note that this approximation also excludes movement tangential to the hypersurfaces:
Without the possibility for summands to cancel each other,
$\text{d}/\text{d}t\,\big < \vec{\sysvec},\vec{\memvec}^{m'}\big >=0$ 
is only fulfilled if $\dot{\vec{\sysvec}}=0$, 
so all remaining solutions are fixed points. Then the following inequalities 
must hold $\forall i$ and $\forall \vec{\sysvec}$ on the surface:
\begin{align*}
 \underbrace{(1-\sysvec_i^2)}_{\geq 0}\memvec_i^{m'}
 \left ( \sum_{m=1}^M   \memvec_i^m \left <  \vec{\memvec}^m,\vec{\sysvec} \right > 
 - \frac{M}{2}\sysvec_i \right )
 &\geq 0 \\
 \memvec_i^{m'}
	\sum\limits_{m=1}^M \memvec_i^{m}
	\left < \vec{\sysvec},\vec{\memvec}^m\right > - \frac{M}{2}\memvec_i^{m'}\sysvec_i &\geq 0
\end{align*}
\begin{equation*}
\big < \vec{\sysvec},\vec{\memvec}^{m'}\big >  \geq
	-\sum\limits_{	\begin{subarray}{l}	
					m=1\\
					m\neq m'
				\end{subarray}
				}^M 
				 \memvec_i^{m} \memvec_i^{m'}
	\left < \vec{\sysvec},\vec{\memvec}^{m}\right > +\frac{M}{2}\memvec_i^{m'}\sysvec_i
\end{equation*}
%

As the left hand side is constant on a hypersurface, the criterion needs 
to be evaluated for a maximized right hand side only and 
 the criterion for the surface can be reduced to one single inequality: 
\begin{align*}
 \big < \vec{\sysvec},\vec{\memvec}^{m'}\big >  
&\geq \max_{i,\vec{\sysvec}}\Bigg (
	-\sum\limits_{	\begin{subarray}{l}	
					m=1\\
					m\neq m'
				\end{subarray}
				}^M 
				 \memvec_i^{m} \memvec_i^{m'}
	\left < \vec{\sysvec},\vec{\memvec}^{m}\right > +\frac{M}{2}\memvec_i^{m'}\sysvec_i\Bigg )\\	
\end{align*}
The sum is maximal 
in $i$ for $\memvec_i^{m}=-\memvec_i^{m'}\,\text{sgn}
(\langle \vec{\sysvec},\vec{\memvec}^{m}\rangle)$ $\forall m \neq m'$, as all scalar products add up.
(If such an $i$ always exists is not relevant here, as we look for a
worst case approximation independent of 
the $\vec{\memvec}^m$.)
The second term is generally much smaller, but $M/2$ at most:
\begin{align*}
&\max_{i,\vec{\sysvec}}
\Bigg (
-\summneqn 
\memvec_i^{m} \memvec_i^{m'}
	\left < \vec{\sysvec},\vec{\memvec}^{m}\right > 
+\frac{M}{2}\memvec_i^{m'}\sysvec_i\Bigg )
\\
\leq &\max_{\vec{\sysvec}}\Bigg (
	\summneqn 
	\left | \left < \vec{\sysvec},\vec{\memvec}^{m}\right > \right | +\frac{M}{2}\Bigg )
\end{align*}
As the maximum of one single $\left | \left < \vec{\sysvec},\vec{\memvec}^{m}\right > \right |$
is much easier to calculate, we approximate an upper bound:
\begin{align*}
\max_{\vec{\sysvec}}\Bigg (
	\summneqn
	\left | \left < \vec{\sysvec},\vec{\memvec}^{m}\right > \right | +\frac{M}{2}\Bigg )\\ 
	\leq 
	\summneqn
	 \max_{\vec{\sysvec}}\left (\left | \left < \vec{\sysvec},\vec{\memvec}^{m}\right > \right | \right )	+\frac{M}{2}	
\end{align*}
In total, our criterion on the hypersurface has reduced to
\begin{align}
C=\big < \vec{\sysvec},\vec{\memvec}^{m'}\big > &
\newtext{\geq} \sum^M_{
\begin{subarray}{l}
m=1\\
m\neq m'
\end{subarray}}
\limits \max_{\vec{\sysvec}}
\left ( \left |\left <\vec{\sysvec},\vec{\memvec}^{m}\right > \right | \right ) +\frac{M}{2}.
\label{eq:hypersurface_crit}
\end{align}
While any hyperplane that fulfills Eq. \eqref{eq:hypersurface_crit} 
confines the system state to larger projections, it is still not trivial to 
evaluate due to the direct dependence on $\vec{\sysvec}$.
\subsubsection{Removing direct dependence on $\vec{\sysvec}$}
$\max_{\vec{\sysvec}}\left ( \left | 
\left < \vec{\sysvec},\vec{\memvec}^{m}\right > \right |\right )$ 
can be approximated as a function of 
$\langle \vec{\sysvec},\vec{\memvec}^{m'}\rangle=C$ 
and inter-pattern scalar products.

First, $\langle \vec{\sysvec},\vec{\memvec}^{m'}\rangle$ is expressed 
with the difference vector 
$\Delta\vec{\sysvec}=\vec{\sysvec}-\vec{\memvec}^{m'}$ between 
 $\vec{\sysvec}$ and 
the closest memorized pattern $\vec{\memvec}^{m'}$:  
\begin{align*}
	\big < \vec{\sysvec},\vec{\memvec}^{m'}\big> 
	&= \big < \vec{\sysvec}-\vec{\memvec}^{m'},\vec{\memvec}^{m'}\big> + \big < \vec{\memvec}^{m'},\vec{\memvec}^{m'}\big>\\
	&= \big < \Delta\vec{\sysvec},\vec{\memvec}^{m'}\big> + N \\ 
	&=\sum_{i=1}^N \limits \Delta\sysvec_i\memvec^{m'}_i  +N 
\end{align*}
 With $\,\text{sgn}(\Delta\sysvec_i)=\,\text{sgn\bf (}\memvec^{m'}_i(\underbrace{\sysvec_i\memvec^{m'}_i}_{\leq 1}-1)\text{\bf )}=-\memvec^{m'}_i$ 
 we get:
\begin{equation}
	\big < \vec{\sysvec},\vec{\memvec}^{m'}\big > = N
	 - \sum_{i=1}^N \limits\left |\Delta\sysvec_i\right| \label{eq:projection_difference_relation}  
\end{equation}
\begin{align*}
\Rightarrow \,\max_{\vec{\sysvec}} \left ( \big | \left < \vec{\sysvec},\vec{\memvec}^{m}\right > \big | \right )
	&= \max_{\Delta\vec{\sysvec}}\left ( \big |\big < \Delta\vec{\sysvec},\vec{\memvec}^{m}\big > 
	+ \big < \vec{\memvec}^{m'},\vec{\memvec}^{m}\big >\big |\right )\\
	&<\max_{\Delta\vec{\sysvec}} \Big( \sum_i \Delta\sysvec_i\memvec^{m}_i \Big ) 
	+ \big |\big < \vec{\memvec}^{m'},\vec{\memvec}^{m}\big > \big | \\
	&=\sum_i \big | \Delta\sysvec_i \big |  
	+ \big | \big < \vec{\memvec}^{m'},\vec{\memvec}^{m}\big > \big |\\
	&=N - \big < \vec{\sysvec},\vec{\memvec}^{m'}\big > + \big | \big < \vec{\memvec}^{m'},\vec{\memvec}^{m}\big > \big |
\end{align*}
\subsubsection{Volumes of growing projection}
Finally, we can remove all direct dependence on $\vec{\sysvec}$ 
from  
 Eq. \eqref{eq:hypersurface_crit}:
\begin{align}
\big < \vec{\sysvec},\vec{\memvec}^{m'}\big >  
&\geq \sum\limits_{
		\begin{subarray}{l}
			m=1\\
			m\neq m'
		\end{subarray}}^M  
	 \max_{\vec{\sysvec}}\left 
	 ( \big | \big < \vec{\sysvec},\vec{\memvec}^{m}\big > \big | \right )	
	 +\frac{M}{2}\nonumber \\	
\big < \vec{\sysvec},\vec{\memvec}^{m'}\big >   &\geq \sum\limits_{
		\begin{subarray}{l}
			m=1\\
			m\neq m'
		\end{subarray}}^M  
	\left ( N - \big < \vec{\sysvec},\vec{\memvec}^{m'}\big >   + \big |\big < 
	\vec{\memvec}^{m'},\vec{\memvec}^{m}
	\big >\big | \right ) +\frac{M}{2} \nonumber \\
M\cdot \big < \vec{\sysvec},\vec{\memvec}^{m'}\big >   &\geq (M-1)\cdot N
+ \sum\limits_{
\begin{subarray}{l}
m=1\\
m\neq m'
\end{subarray}}^M  \big |\big < \vec{\memvec}^{m'},\vec{\memvec}^{m}\big > \big |
+\frac{M}{2} \nonumber 
\end{align}
\begin{equation}
\big < \vec{\sysvec},\vec{\memvec}^{m'}\big >   \geq \frac{M-1}{M}\cdot N
+ \frac{1}{M}\sum\limits_{
\begin{subarray}{l}
m=1\\
m\neq m'
\end{subarray}}^M  \big |\big < \vec{\memvec}^{m'},\vec{\memvec}^{m}\big > \big |
+\frac{1}{2} 
\label{eq:final_surface_crit}
\end{equation}
This final criterion for a confining hyperplane does 
not depend on a point on the surface.
 
Additionally, every surface $\big < \vec{\sysvec},\vec{\memvec}^{m'}\big >= C_{min} $  that fulfills Eq. \eqref{eq:final_surface_crit} defines a {\it volume of growing projection} 
for larger $C$: As the right hand side 
of Eq. \eqref{eq:final_surface_crit} is constant, all hyperplanes with $C>C_{min}$ fulfill the criterion as well.

If several attractors existed in the confined space, however, 
no conclusion could be made on the basins of attraction, as a confined system state
 could move to any of them.
Therefore, we exclude that any attractor besides $\vec{\memvec}^{m'}$ 
exists in a volume of growing projection:
 
\subsubsection{${\vec{\memvec}^{m'}
\text{ being the only attractor enclosed}}$}
Assume an attractor $\vec{\memvec}^a$ exists 
inside the region defined by Eq. \eqref{eq:final_surface_crit}. 
Now consider a small perturbation around $\vec{\memvec}^a$
that increases $\langle \vec{\sysvec}, \vec{\memvec}^{m'} \rangle$,
 for example 
  $\epsilon \memvec_i^{m'}\cdot \hat{e}_i$ 
 if $\sysvec_i^a \neq \memvec_i^{m'}$.
As 
$\text{d}/\text{d}t\langle \vec{\sysvec},\vec{\memvec}^{m'}\rangle  \geq 0 $
 in the confined space,
the system cannot relax back to $\vec{\memvec}^a$.
No non-isolated attractor exists (see Sec. \ref{sec:fixed_points}), 
so $\vec{\memvec}^a$ has at least one unstable eigendirection which 
contradicts the assumption that $\vec{\memvec}^a$ is an attractor.

The only exception is the attractor $\vec{\memvec}^{m'}$ itself: As it has the largest 
projection on itself, all perturbations must lower
 $\langle \vec{\sysvec},\vec{\memvec}^{m'}\rangle$.\\

Summing up: Every system state $\vec{\sysvec}$ that 
obeys Eq. \eqref{eq:final_surface_crit} 
 must be in the basin of attraction of $\vec{\memvec}^{m'}$,
as projection on $\vec{\memvec}^{m'}$ increases monotonically  along the trajectory 
and $\vec{\memvec}^{m'}$ is the only attractor for larger projections.

\subsection{Guaranteed recognition}
\subsubsection{Recognition criteria}
As any defective initialized pattern is binary, it can be characterized 
by the number of defective pixels $\nfalse$ in which defective input pattern 
and correct memorized 
pattern are different. 
Eq. \eqref{eq:final_surface_crit} can be solved 
for $\nfalse$ with Eq. \eqref{eq:projection_difference_relation}, as 
$\nfalse$ is a special case of 
$\sum_i\left |\Delta\sysvec_i\right|/2$:

\begin{align}
N - 2\nfalse   &> \frac{M-1}{M}\cdot N
+ \frac{1}{M}\sum\limits_{
\begin{subarray}{l}
m=1\\
m\neq m'
\end{subarray}}^M  \big |\big < \vec{\memvec}^{m'},\vec{\memvec}^{m}\big > \big |
+\frac{1}{2} \nonumber \\
\nfalse &< \frac{1}{2M}\bigg (N-
\sum_{
\begin{subarray}{l}
m=1\\
m\neq m'
\end{subarray}}^M  \big |\big < \vec{\memvec}^{m'},\vec{\memvec}^{m}\big > \big |\bigg )-\frac{1}{4}
\label{eq:final_general_crit}
\end{align}
(The equality in Eq. \eqref{eq:final_surface_crit} must be dropped here, 
as perturbations and higher order terms neglected 
in Eq. \eqref{eq:theta_dynamics} might push a defective pattern on the outermost hyperplane out of the confined space.)\\
For {\it pairwise orthogonal patterns}, 
$\langle \vec{\memvec}^{m'},\vec{\memvec}^{m}\rangle =0$ $\forall m\neq m'$ and Eq. \eqref{eq:final_general_crit} becomes:
\begin{equation}
\boxed{\nfalse < \frac{N}{2 M}-\frac{1}{4}} \label{eq:orthogonal_matching_criterion}
\end{equation}
We now treat 
{\it general patterns} with $\langle \vec{\memvec}^{m'},\vec{\memvec}^{m}\rangle \neq0$.
A criterion that does not 
depend on the correct memorized pattern $\vec{\sysvec}^{m'}$ is obtained with 
the definition 
$ \Sigma_{max}=\max_{\vec{\sysvec}^{\tilde{m}}}
	( 
	\sum_{m=1,m\neq \tilde{m}}^M  | \langle \vec{\memvec}^{\tilde{m}},\vec{\memvec}^{m}\rangle 
	 | )>\sum_{m=1,m\neq m'}^M | \langle \vec{\memvec}^{m'},\vec{\memvec}^{m}\rangle |
$
from Sec. \ref{sec:fixed_points}. Then the worst case of Eq. \eqref{eq:final_general_crit} is 
\begin{equation}
\boxed{\nfalse < \frac{N-\Sigma_{max}}{2 M}-\frac{1}{4}} \label{eq:general_matching_criterion}.
\end{equation}
Eq. \eqref{eq:general_matching_criterion} guarantees successful recognition for arbitrary patterns.
\subsubsection{Consistency Check}
The basin of attraction has to vanish when the fixed point looses stability. 
Therefore, we can regain stability criteria for the $\vec{\memvec}^{m}$ 
by minimizing the necessary extension of the basin of attraction
 in Eq. \eqref{eq:orthogonal_matching_criterion} and Eq. \eqref{eq:general_matching_criterion},
which corresponds to $\lim_{\nfalse \rightarrow 0}\limits$:
  \newtext{\begin{align*}  
  \lim_{\nfalse \rightarrow 0} \limits \text{Eq. \eqref{eq:orthogonal_matching_criterion}}: \quad0&<\frac{N}{2M} -\frac{1}{4} \\
  M&< 2 N
 \end{align*}
 }
This coincides with our calculation that pairwise orthogonal patterns are always stable:
At most, $N$ orthogonal patterns can exist, as they are linear independent and 
$\dim(\text{span}(\{\vec{\memvec}^m\}))\leq N$, so $M < 2 N$ is always fulfilled.
 \newtext{\begin{align*}  
  \lim_{\nfalse \rightarrow 0} \limits\text{Eq. \eqref{eq:general_matching_criterion}}: \quad0&<\frac{N-\Sigma_{max}}{2M} -\frac{1}{4} \\
  \Sigma_{max}&< N-\frac{M}{2}
 \end{align*}
 }
This again reproduces our result for the 
stability of non-orthogonal patterns.
\section{Numerical Simulations}
\label{sec:sim_details}
In this section we validate our criterion for successful pattern 
recognition with simulations of the full phase dynamics
 Eq. \eqref{eq:phase_equations}.
\subsection{Numerical methods and parameters}
The equations have been implemented in C 
and integration was performed with the classical Runge-Kutta method.
A timestep $dt=1\cdot 10^{-4}$ and
a coupling strength $\epsilon=0.1$ were used. The angular frequencies 
were distributed 
according to $\Omega_i=1200 + 1800\cdot G_i/G_N$, 
where $G_i$ is 
the i-th element of a Golomb ruler \cite{golomb1997} 
(see also Appendix \ref{app:averaging}).
The near optimal Golomb rulers used  were both 
taken from \cite{atkinson1984_golomb_examples}:
 $\{$0, 17,  20,  86,  119,  140,  166,  227,  240,  255,  
353,  430,  520,  559,  564,  565,  602,  675,  724, 781, 
817, 833, 905, 929, 961, 970, 980, 1131, 1162, 1189, 
1212, 1319, 1403, 1433, 1437, 1451, 1462, 1497, 1504, 1589, 
1601, 1680, 1763, 1785, 1825, 1880, 1888, 1956, 1958$\}$ for $N=49$  
and 
$\{$0, 34, 44, 91, 95, 147, 207, 278, 332, 364, 
	375, 405, 458, 520, 682, 698, 701, 710, 853, 868, 
	901, 946, 973, 1022, 1080, 1150, 1155, 1172, 1240, 1254, 
	1290, 1429, 1540, 1546, 1605, 1642, 1682, 1684, 1705, 1751, 
	1771, 1806, 1835, 1943, 1967, 2041, 2151, 2164, 2182, 2189, 
	2190, 2270$\}$ for $N=52$.
%
%

For simulations in Fig. \ref{fig:successful_simulation} 
and Fig. \ref{fig:failing_simulation}, defective patterns 
were chosen manually and 
memorized patterns are taken from Fig. \ref{fig:output_patterns}. 
All pseudorandom numbers (necessary for random distribution of erroneous pixels and 
construction of random orthogonal patterns) were created using C's standard random number generator rand() from stdlib, 
which was seeded with the time in microseconds times the process ID.
\subsection{Testing criteria for guaranteed recognition}
\label{sec:SimTesting}
In order to test criteria Eq. \eqref{eq:orthogonal_matching_criterion} 
and \eqref{eq:general_matching_criterion}, simulations were performed for both the 
non-orthogonal patterns shown in Fig. \ref{fig:output_patterns} 
with N=49 pixels as well as
for 3 random orthogonal patterns with N=52 pixels. 
Simulations started after setting the initial conditions to a defective pattern 
similar to one of the memorized patterns but different in exactly $\nfalse$
 randomly distributed erroneous pixels. 
 In order to save simulation time, simulations were aborted if the 
 system state reached one of the memorized patterns, as they are proven 
 to be attractors. In all other cases, simulations were continued until 
 $\vert\sysvec_i\vert\geq 0.9\,\forall i$ for a period $t_{wait}=500$.
Recognition success was 
tested by projecting the $\vec{\sysvec}$-coordinates 
of the final system state on the memorized patterns: 
If $\langle\vec{\sysvec},\vec{\memvec}^{m'}\rangle /N > 0.99$, recognitions 
were counted as successful.

For the non-orthogonal patterns with $M=3$, $N=49$ and $\Sigma_{max}=10$, 
the recognition criterion Eq. \eqref{eq:general_matching_criterion} predicts 
recognition success for $\nfalse<(N-\Sigma_{max})/(2M)-0.25=6.25$. 
300 simulations were performed for $\nfalse\in\{6..16\}$ for each pattern 
and results are summed up in Table \ref{tab:table_N49}.
\begin{table}
\includegraphics{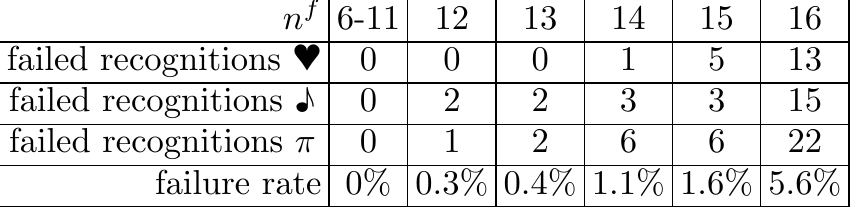}
\caption{Failed Recognitions with non-orthogonal patterns as shown in Fig. \ref{fig:output_patterns}. 300 
recognitions were performed for each pattern and each number of erroneous pixels $\nfalse$. 
Erroneous pixels were distributed randomly for each simulation.
\label{tab:table_N49}}
\end{table}
All recognitions were successful for $\nfalse \leq 11$ and 
the rate of failed recognitions grows slowly for larger $\nfalse$. 
Obviously, our criterion seems to be too strict.

Similarly, 1000 simulations were performed with  
orthogonal random patterns with $N=52$ 
and  
$M=3$
 for each $\nfalse\in\{8..17\}$. Here,  $\nfalse<N/(2M)-0.25=8.42$ 
 is predicted by Eq. \eqref{eq:orthogonal_matching_criterion}.
 Random orthogonal patterns were constructed 
 by using the elementwise product $\circ$: 
 As orthogonal patterns with $\memvec_i\in\pm1$ differ in exactly 
 $N/2$ pixels, a pattern $\vec{\memvec}^2$ orthogonal to any pattern 
 $\vec{\memvec}^1$ can be easily found by creating a
 ''difference vector'' $\vec{d}^{1,2}$, where $N/2$ $+1$- 
 and $-1$-entries are randomly distributed.
 Then $\vec{\memvec}^2=\vec{\memvec}^1 \circ \vec{d}^{1,2}$. 
 
 For 3 orthogonal patterns,
 $\vec{\memvec}^1$, $\vec{d}^{1,2}$ and $\vec{d}^{1,3}$ were first chosen randomly.
  Then
 $\vert\langle \vec{\memvec}^2,\vec{\memvec}^3  \rangle \vert = 
 \vert\langle \vec{d}^{1,2},\vec{d}^{1,3}  \rangle \vert$ was minimized 
 by switching 2 randomly selected pixels in a randomly selected difference 
 vector, if the absolute value of the scalar product diminished.

Results are summed up in Table \ref{tab:table_N52}. 
 \begin{table}
\includegraphics{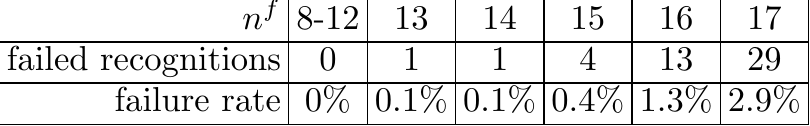}
\caption{Failed Recognitions with random orthogonal patterns with $N=52$ pixels. 
1000 recognitions were performed for each number of erroneous pixels $\nfalse$.
 Random distribution of erroneous pixels and the construction of 
 random orthogonal patterns was repeated for each simulation.
 \label{tab:table_N52}}
\end{table}
Similar to the simulations with non-orthogonal patterns, 
 \newertext{recognitions are always successful for $\nfalse\leq 12$, which is
  significantly larger than} 
 predicted 
by the criterion for guaranteed recognition.  
\newertext{For even larger $\nfalse$, the rate of failed recognitions stays small.}
\subsection{Failed recognitions are rare events}
One might expect that the criterion for guaranteed recognition is not 
optimal for both the orthogonal random patterns and 
our choice of non-orthogonal patterns, so that 7 respectively 9  
erroneous pixels  or even more can always be correctly recognized as well.
However, failed recognitions are just rare for $n^f=7$ / $n^f=9$ instead. 
We now construct problematic starting patterns with $n^f=7$ for the non-orthogonal memorized patterns 
that fail
 in the recognition process:

According to Eq. \eqref{eq:hypersurface_crit}, recognition will fail 
 if the scalar products between the defective starting pattern 
and non-similar memorized patterns are extremized. Considering the scalar products 
 $\left < \vec{\memvec}^{\varheart}, \vec{\memvec}^{\pi}\right >=+5$, 
$\left < \vec{\memvec}^{\varheart},\vec{\memvec}^{\text{\textmusicalnote}}\right >=-5$,
and $\left < \vec{\memvec}^{\pi},\vec{\memvec}^{\text{\textmusicalnote}}\right >=-1$, an erroneous 
heart-pattern is most likely to fail.
Assume furthermore that the number of erroneous pixels $\nfalse$ is fixed. 
Then the right hand side of
 Eq. \eqref{eq:hypersurface_crit} can be maximized by distributing
 the errors on positions where they increase the projection on the $\pi$- and decrease the projection on the 
$\text{\textmusicalnote}$-pattern.
10 such ''worst-case'' positions can be found for the $\varheart$-pattern and  
$\left (\begin{smallmatrix} 10 \\ 7 \end{smallmatrix}\right ) =120$ possible combinations exist to 
 distribute $\nfalse=7$ erroneous pixels on the ''worst-case'' positions.

\begin{figure}[h!]
\centering
	\includegraphics{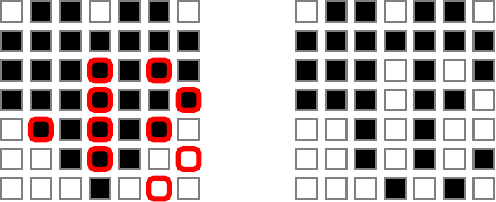}
	\caption{
On the left side, an unperturbed $\varheart$-pattern is shown.
Erroneous pixels on 
red-circled positions extremize the sum of inter-pattern scalar products. 
All erroneous $\varheart$-patterns with 7 erroneous pixels on marked locations fail the recognition process.
Simulations of all such patterns resulted in the spurious attractor 
shown on the right.	
}
\label{fig:worstCasePattern}
\end{figure}

Simulations were performed for all of these ''worst case patterns''.  Recognition failed for all simulations 
and the system state relaxed to an attractor with projections of 
0.59, -0.51  and 0.51
on the $\varheart$-, $\text{\textmusicalnote}$-  and $\pi$-pattern.
Possible worst-case positions for erroneous pixels 
and the irregular output pattern 
are shown in
 Fig. \ref{fig:worstCasePattern}.
Indeed, simulations with randomly distributed errors could not recognize this: As there are 
$\left (\begin{smallmatrix} 49 \\ 7 \end{smallmatrix}\right ) \approx 8,6 \cdot 10^7$ possibilities
 to distribute the erroneous pixel on the pattern and only 
 $\left (\begin{smallmatrix} 10 \\ 7 \end{smallmatrix}\right ) =120$ worst case distributions 
 can be found, the chance to encounter a failing random starting pattern 
 is almost negligible. Furthermore, all $\left (\begin{smallmatrix} 10 \\ 6 \end{smallmatrix}\right ) =210$ worst-case-patterns 
 for $\nfalse=6$ were successfully recognized as the $\varheart$-pattern in simulations, 
 which again validates Eq. \eqref{eq:general_matching_criterion} as criterion for guaranteed recognition.
 Similar calculations can be performed for the 
 orthogonal case. This is a good example 
 that extracting basins of attractions in high-dimensional systems 
 with simulations can only give an approximation on the success rate 
 but no guaranteed criterion. 
 From another point of view, failed recognitions are rare, so a higher $n^f$
 is acceptable if a non-perfect recognition rate is sufficient.
\section{Discussion}
\label{sec:discussion}
MONACO gains its distinctive properties from two design features:
 \begin{enumerate}
	\item \emph{Two mirrored globally coupled subnetworks} are used. \\
		First of all, the use of two groups enables the internal 
		generation of the coupling modulations. 
		Second, the effective coordinates of the network are phase 
		differences $\deltheta_i$ of oscillators of equal frequency.
		Values of phase differences can easily be read out by 
		multiplying signals of an oscillator pair and 
		using a low-pass filter, gaining $\cos{\deltheta_i}$. Similarly, 
		setting the 
		initial conditions requires only positive or negative coupling 
		between two oscillators forming a pair.
		Third, the effective average coupling strength $\epsilon$
		 is doubled with two subnetworks, enabling faster recognition 
		  (compare with Appendix \ref{app:averaging}).
		  The fourth advantage is much subtle: In all architectures 
		 with externally generated {\cf{}}s, frequencies in the \cf{} are fixed to the
		 natural frequencies of the uncoupled oscillators. 
		 However, the so-called ''acceleration effect'' \cite{Aonishi2002_AccelerationEffect} changes 
		 the frequencies of even weakly coupled oscillators.
		  \newertext{Any} mismatches 
		  between oscillator frequency and \cf{} frequency component\newertext{s 
		   w}ould further limit the coupling strength $\epsilon$.
		  As a higher coupling strength reduces recognition time, we decided to avoid the problem altogether: 
		  Since oscillators in both networks are affected symmetrically by the coupling, the acceleration effect 
		  will be equal and frequencies in the {\cf{}}s are adjusted \newtext{automatically.}
		  It is noteworthy, that the coupling between single oscillator 
		  pairs is above the Kuramoto threshold and thus frequencies of 
		  the two oscillators adapt. Hence, the architecture allows 
		  for some tolerance in the frequency mismatch of an oscillator 
		  pair.
	\item Novel \emph{coupling modulations} are used.\\
		As shown in Eq. \eqref{eq:simplified_coupling_modulations}, 
		the used \cf{}s can be constructed 
		with $\mathcal{O}(N\cdot M)$ connections only. 
		Note that there cannot be any better scaling, 
		as patterns consist of $N\cdot M$ independent pixels. 
		Additionally, this {\cf{}}s introduce novel effective dynamics 
		Eq. \eqref{eq:theta_dynamics}, where the only existing attractors 
		are isolated fixed points with $\cos{\deltheta_i}\in\{\pm 1\}$ 
		(Section \ref{sec:fixed_points}).
		As every pixel settles at these binary values, 
		the output is inherently digital, 
		which further simplifies readout and subsequent processing.
		 All memorized patterns are attractive if inter-pattern 
		scalar products are not too large (see Eq. \eqref{eq:general_stability_criterion} for guaranteed stability).
		As memorized patterns are no transient phenomenon, but long-term stable, 
		readout does not need to be exactly timed and the output can be retrieved at a later time.
		 Furthermore, the dynamics allow us to calculate a lower bound on the 
		basins of attraction analytically (Section \ref{sec:boa_a_matching_crit}).
		This leads to a non-probabilistic criterion for guaranteed recognition 
		that includes finite-size effects, Eq. \eqref{eq:general_matching_criterion}.
 \end{enumerate}
Note that the mirrored subnetwork structure should not be confused with ''layers'' from
 ''traditional'' layered  neural networks. 
MONACO is very similar to a
 continuous version of the Hopfield model\cite{Hopfield1982_original} (compare with Eq. \eqref{eq:alpha_dynamics} ):
Each oscillator pair corresponds to an artificial neuron that 
''stores'' its phase difference $\deltheta_i$. 
The synchronization process can be seen as 
continuous updating of the $\deltheta_i$. 
MONACO's subnetworks, however, change the properties of the ''neurons'',  
while the ideal effective dynamics Eq. \eqref{eq:theta_dynamics} 
 remain unchanged except for the coordinates they are represented in. This is distinct from more ''traditional'' 
layered neural networks, where the layer structure is essential to the 
dynamics.

On the contrary, a design with two subnetworks is not necessary in order to 
obtain the described dynamics including isolated attractors: 
A multiplicative \cf{} suffices; consider e.g. the following single-network-system:
\begin{align*}
\dot{\vartheta}_i &=\Omega_i+\cos{\vartheta_i} \cdot
	a_{ext}(t)\cdot\frac{\epsilon}{N}\sum_{j=1}^N\limits \sin{\vartheta_j} \\
{a}_{ext}(t)&=\sum_{k,l=1}^N\limits S_{kl} \sin{\Omega_kt}\sin{\Omega_lt}
\end{align*}
Here, the averaged dynamics would be the same as for MONACO, but in 
coordinates $\varphi_i(t)=\vartheta_i(t)-\Omega_i t$ (compare with Appendix \ref{app:averaging}):
\begin{equation*}
	\dot{\varphi}_i = - \frac{\epsilon}{2N}\left ( \sum_{j=1}^N\limits \cmatrix_{ij} 
	\sin{\varphi_i}\cos{\varphi_j}-\frac{M}{2}\sin{\varphi_i}\cos{\varphi_i}\right )
\end{equation*}
Phase shifts $\varphi_i$
must be used, as no oscillators with equal frequencies exist in this setup and therefore,  
phase differences $\deltheta_i$ are no useful coordinate. As discussed below, 
tracking changes of the $\varphi_i$ 
requires very precise frequency and time measurements,
which renders readout difficult and error-prone. 
Therefore, this exemplary network is inferior to MONACO.

\begin{figure*}
	\begin{subfigure}[b]{0.3\textwidth}
		\parbox[c][0.87\textwidth][c]{\textwidth}{\centering\includegraphics[scale=0.66]{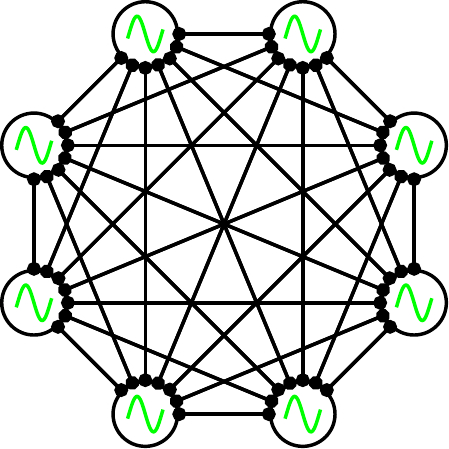}}
		\caption{all-to-all connected\\ networks (I) and (III)}
	\end{subfigure}
	\begin{subfigure}[b]{0.3\textwidth}
		\parbox[c][0.87\textwidth][c]{\textwidth}{\centering\includegraphics[scale=0.66]{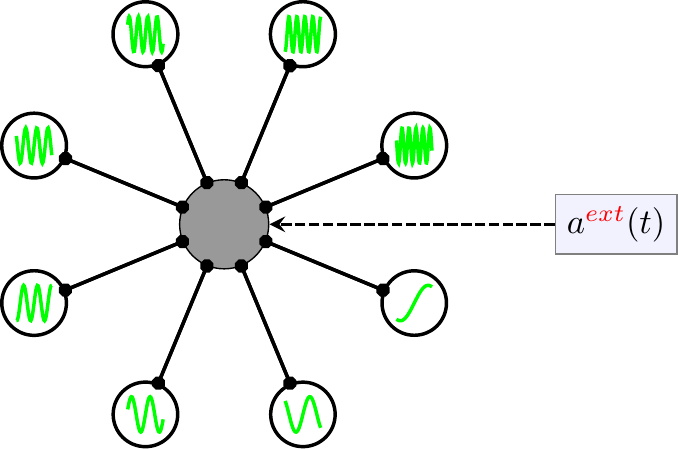}}
		\caption{dynamically all-to-all connected (IIA)}
	\end{subfigure}
	\begin{subfigure}[b]{0.3\textwidth}
		\parbox[c][0.87\textwidth][c]{\textwidth}{\centering\includegraphics[scale=0.66]{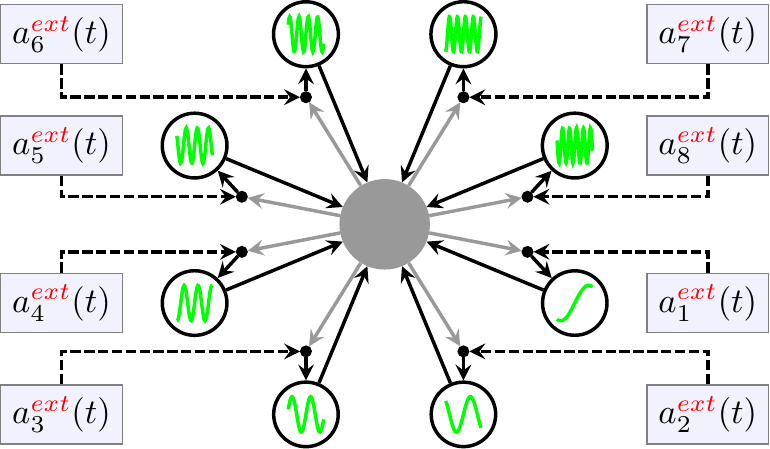}}
		\caption{dynamically all-to-all connected (IIB)}
	\end{subfigure}
	\caption{Schematics of previous oscillatory neural network architectures that act as autoassociative memories
	\label{fig:schematics_comparison}}
\end{figure*}

The
 MONACO-architecture will now be compared to other associative memories 
 consisting of phase oscillators. Distinctive features are 
 compared in Table \ref{tab:features_comparison}, while schematics of 
  are shown in Fig. 
 \ref{fig:schematics_comparison}. 
\begin{table*}
	\includegraphics[width=\linewidth]{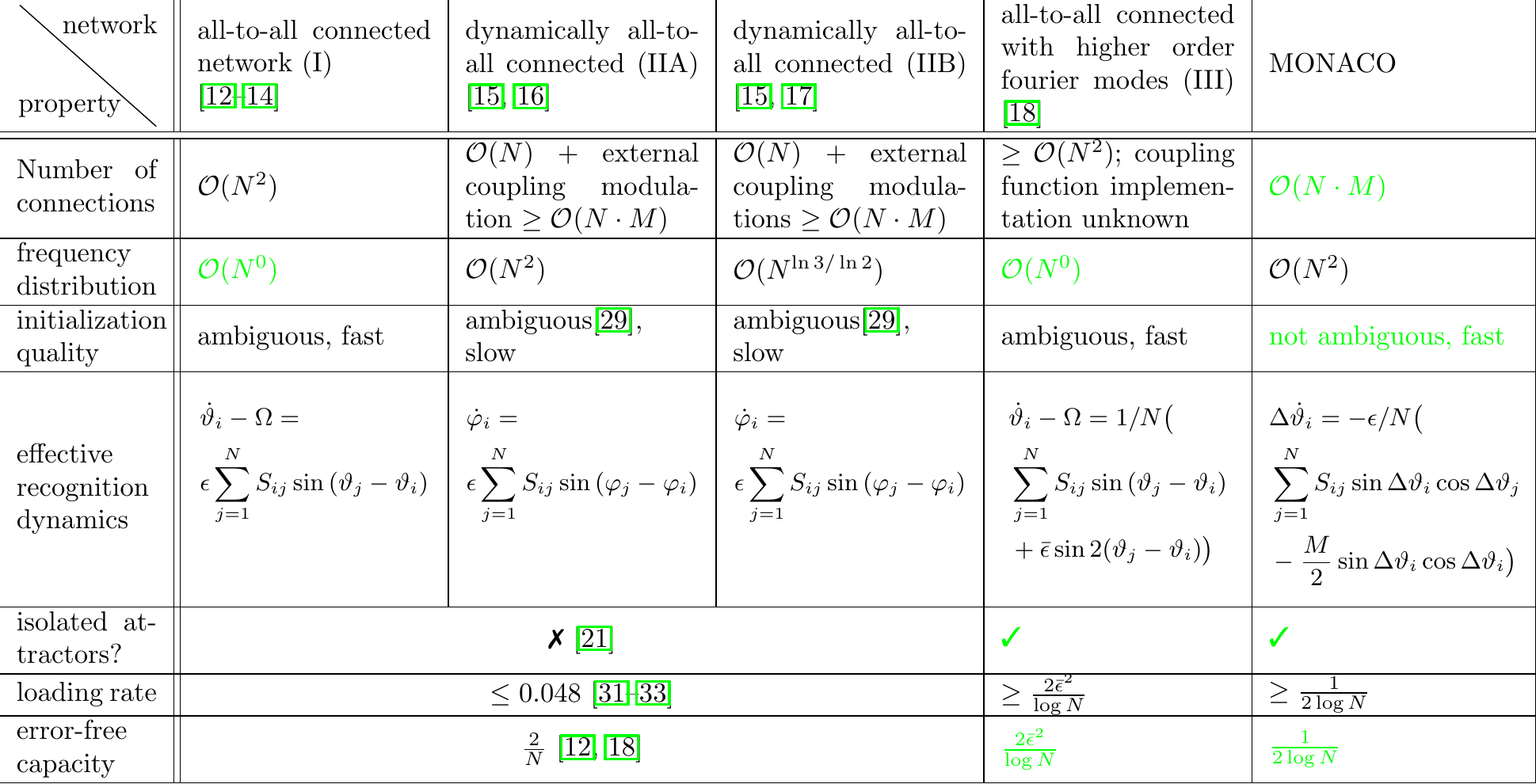}
	\caption{Comparison of MONACO with other autoassociative memory architectures based on phase-oscillator networks. The best 
	performances for every property are marked in green.
	\label{tab:features_comparison}
	}
\end{table*}

 We discriminate between two types of networks: In \emph{physically} 
 all-to-all connected networks (architectures 
 (I)\cite{aonishi1998_phase_transitions,aoyagi_1997_cutting_connections,arenas_1994_longterm}  
 and 
 (III)\cite{nishikawa2004_second_order}), oscillators 
 have the same frequencies and every oscillator is connected with
  every other(see Fig. \ref{fig:schematics_comparison} a). 
  Therefore, the number of connections scales with 
  $\mathcal{O}(N^2)$ in these networks, which limits the networks' 
  size. 
  As proposed in \cite{Hoppensteadt1999}, oscillators of different 
  frequency can be all-to-all connected \emph{dynamically}  
  with only one physical connection per oscillator if the oscillators' 
  coupling is modulated in time. 
  In architecture (IIA)\cite{Hoppensteadt1999,Hoelzel2011}, the oscillators are globally coupled to 
  a sum of the oscillators' signals with a single temporal modulation of the coupling
  (see Fig. \ref{fig:schematics_comparison} b). Due to the global coupling, 
  the number of connections scales with $\mathcal{O}(N)$ connections only.
  
  Architecture (IIB)\cite{Hoppensteadt1999,Kostorz2013} follows a slightly more complicated scheme, where 
  every oscillator receives the signals of all other oscillators, but 
  each oscillator has its own coupling modulation 
  (see Fig. \ref{fig:schematics_comparison} c). Nevertheless, 
  the scaling of the number of 
  connections is still $\mathcal{O}(N)$.

  MONACO is a dynamically all-to-all connected network as well. 
  Each subnetwork is globally coupled similar to (IIA), 
  albeit with a different coupling modulation (see Fig. \ref{fig:crosswise}). 
  The use of two mirrored subnetworks allows for the
 {\it internal} generation of the global coupling modulations.
  In contrast, the hardware implementation 
  for architectures (IIA) and (IIB) introduced
  in \cite{Hoelzel2011,Kostorz2013} was fed by 
   computer-generated coupling modulations. 
   In MONACO, the scaling of the number of 
  connections is $\mathcal{O}(N\cdot M)$
  (see Eq. \ref{eq:simplified_coupling_modulations} for the coupling 
  modulations and consider that global 
  coupling scales with $\mathcal{O}(N)$). 
  This scaling is optimal if the 
  generation of the coupling modulations is considered, 
  as $N\cdot M$ pixels have to be incorporated.

 However, the reduction in the number of spatial connections
 is not for free: The original complexity in space is transferred
 to a complexity in time with the number of frequencies contained
 in the coupling modulation growing like $\mathcal{O}(N^2)$ 
 for architecture (IIA) and MONACO\cite{Hoppensteadt1999}. Frequency conditions for 
 architecture (IIB) are less restrictive and the number of frequencies scales with 
 $\mathcal{O}(N^{\ln{3}/\ln{2}})$\cite{Kostorz2013}.
 
 Now, coordinates of the network dynamics will be discussed, as 
 they determine how initial conditions can be enforced as well as how the 
  system state can be read out. 
 In (I) and (III), the desired dynamics occur in oscillators' phases 
 $\vartheta_i=\Omega t + \varphi_i$, so pixels of the same value have 
 the same phase. An encoded pattern is then represented by two groups of 
 oscillators whose phases differ by $\pi$. Note that this 
 representation itself is ambiguous, as it is physically impossible to decide 
 if a group follows or precedes the other one. In other words, the physical 
 state represents a pattern as well as its inverse. 
 In (IIA) and (IIB), equal pixels are represented 
 by equal phase shifts $\varphi_i$ and different pixels differ by 
 a phase shift difference of $\pi$. 
 Note that phase shifts are only unique up to a constant 
 $\varphi_i^0=\varphi_i(t=0)$. As a consequence, only differences 
 $\varphi_i(t)-\varphi_i(t')$ can be determined.
 In MONACO patterns are coded into phase differences $\deltheta_i=
 \vartheta_i^{[1]}-\vartheta_i^{[2]}$ of oscillators of equal frequencies. 
 Each pixel is mapped onto a phase difference with 
 $\alpha_i=\cos{\deltheta_i}$,
  so $\alpha_i=+1$ corresponds to a synchronized oscillator pair and 
  $\alpha_i=-1$ to a antisynchronized one. Therefore, MONACO's system 
  state represents a pattern without ambiguity.
  
The different nature of the variables entail that 
 also the setting of initial conditions differs radically between architectures:  
 Phase differences in MONACO are easily manipulatable: Oscillator pairs
  corresponding 
 to $+1$ are directly coupled positively, while pixels with $-1$  
  receive a negative coupling, resulting in synchronized pairs with 
   $\deltheta_i=0$ or  $\deltheta_i=\pi$. 
  Phases $\vartheta_i$ change quickly in time, so they are difficult to 
  control directly. However, 
  initial conditions in (I) and (III) can be set similar to MONACO by coupling 
 all $N$ oscillators in a row, where oscillators representing equal
  pixels are coupled positively and 
 unequal pixels interact via a negative coupling. 
 In (IIA) and (IIB), two main problems must be overcome to set initial conditions:
 First, phase shifts cannot be manipulated directly and second, 
 phase shifts are undefined without a temporal reference.  
 Hoppensteadt and Izhikevich\cite{Hoppensteadt1999} proposed to use the 
 same coupling circuitry as used for the recognition, but with a
  different coupling matrix $\matr{\cmatrix}$: 
  $\cmatrix_{ij}=\sysvec_i^d\sysvec_j^d$ is 
 used to initialize a defective pattern $\vec{\sysvec}^d$.
  Then, recognition is performed with the usual coupling matrix 
  $\cmatrix_{ij}=\sum_{m=1}^M \memvec_i^{m}\memvec_j^{m}$. By evaluating 
   phase shift changes between the introduced initial condition and 
   the recognition, pixel changes can be retrieved without 
   the constants $\varphi_i^0$.
   However, initialized patterns are ambiguous: As  
   $\cmatrix_{ij}(-\vec{\memvec}^d)=(-1)^2\memvec_i^d\memvec_j^d
   =\cmatrix_{ij}(\vec{\memvec}^d)$, 
   the inverse pattern $-\vec{\memvec}^d$ is initialized half of the time.
   Additionally, as this method is 
   limited by the averaging condition similar to the recognition,
    this method is of timescale $1/\epsilon$ and therefore considerably 
    slower than the direct coupling used for (I), (III) and MONACO.

 Similarly, readout of the final pattern is easy in MONACO: As mentioned 
 above, $\cos{\deltheta_i}=\sysvec_i$ can be read out directly from 
 the corresponding oscillator pair. Readout in (I) and (III) is analogue, 
 but phase differences between different pixels are determined, which 
 again describes both a specific pattern and its inverse.
 For (IIA) and (IIB), phase shifts have to be determined by comparing 
 the phase of an oscillator with an external reference. Then, the difference 
of phase shifts between final state and the initial conditions needs to be evaluated. 
 In refs. \cite{Hoelzel2011,Kostorz2013}, this was done with a computer and 
 analog-digital converter 
 cards \cite{footnote_readout_manual,footnote_readout_noAmbiguity}.
 
 Ease of readout additionally depends on the effective dynamics 
 of the architectures:
 Traditional Kuramoto-type networks (I)
  employ a coupling that depends only on the mutual phase differences
  of all oscillators ($\propto \sin{(\vartheta_{i}-\vartheta_{j})}$).
  While (IIA) and (IIB) have a seemingly more complex structure
  due to their coupling modulations, dynamics are effectively the same 
  as in (I) 
  after averaging (Compare with Table \ref{tab:features_comparison}), 
  albeit in different coordinates. In these dynamics, the individual
 patterns are not individual attractors, but part of one large attractor. 
 More precisely, patterns are connected by lines of 
 attractive non-isolated fixed points\cite{Hoelzel2015_stability}.
 Consequently, recognition is only possible for short times, 
 as the system state drifts on the attractor 
 due to implementation inaccuracies 
 and higher order terms and readout must occur immediately after the 
 recognition is successful. Additionally, the system state does only 
 settle close to the correct memorized pattern, so the output values 
 are not inherently digital as the patterns are.
  
MONACO's dynamics ({\it cf.} Eq. \eqref{eq:phase_equations}) 
take on a simple mathematical form after averaging 
(Eq. \eqref{eq:theta_dynamics} for the formulation in
 phase differences, Eq. \eqref{eq:alpha_dynamics} in pattern space $\vec{\sysvec}$).
 In these novel dynamics, binary memorized patterns are {\it individual} attractors. 
In \cite{nishikawa2004_second_order}, 
yet another dynamics was introduced with architecture (III) 
(see Table \ref{tab:features_comparison}). Memorized patterns are isolated 
attractors here as well due to higher order Fourier modes in the coupling function.
Due to the isolated attractors,
 readout does not need to be exactly timed and the output is 
 inherently digital in MONACO as well as in architecture (III). 
 Additionally, the dynamics of (III) enable the exclusion of 
 spurious attractors for specific 
 parameter ranges, while MONACO's dynamics allowed us to determine 
 lower bounds on the basins of attraction, as discussed below and in 
 Section \ref{sec:boa_a_matching_crit}.

Concerning quantitative measures for associative networks,
 often the capacity or loading rate of a network is used.
  It describes the maximum possible ratio of $M$
 and $N$, 
 where the system state still settles 
 close to the correct memorized pattern.
 Usually, it is computed for a set of random memorized patterns in the limes 
 $N\rightarrow \infty$.  
 This definition, however, includes deviations from the memorized patterns, 
 so e.g. some bits may be erroneous at retrieval. 
 Nishikawa et al. point out the importance of error-free retrieval for 
 engineering applications \cite{nishikawa2004_second_order} and remind of the 
   error-free capacity (def. in Subsec. \ref{subsec:errorfree}) as a more meaningful quantity, 
    as it is used for traditional neural networks \cite{mceliece1987_capacity}. 
  The error-free capacity of MONACO (Eq. \eqref{eq:errorfree_capacity}) 
  is on a par with architecture 
  (III)\cite{nishikawa2004_second_order} and equal 
  to the error-free capacity of the Hopfield model \cite{mceliece1987_capacity}  
   while memorized patterns are typically unstable in architectures (I), (IIA) 
  and (IIB) with an error-free capacity of 
  $2/N$\cite{nishikawa2004_second_order,aonishi1998_phase_transitions}.
  The loading rate for architectures (I),(IIA) and (IIB) has been derived as  
   $0.048$ \cite{aoyagi1998retrieval,cook1989_meanfield_QState,uchiyama2002stability}, 
   while it has not been calculated for neither architecture (III) nor MONACO yet. 
   However, the error-free capacities are lower bounds on the loading rates and 
   may be larger than the value for  (I) - (IIB) similar to the error-free capacities.

 While the loading rate and the error-free capacity are useful for 
 comparing architectures, their probabilistic nature and the derivation for 
 $\lim N \to \infty$ impair their significance for real networks: 
 Specific sets of memorized patterns are possibly not random  and 
 finite size effects might improve or impair pattern stability 
 as well as recognition success. 
  Non-probabilistic criteria 
 valid for all network sizes allow to exactly evaluate performance of 
 a network for a specific use case and enable the development of more 
 complex algorithms using the recognition process repeatedly. We 
 derived such criteria for MONACO:
 Eq. \eqref{eq:general_stability_criterion} guarantees stability of all 
 memorized patterns if scalar products between memorized patterns 
 are not too large. 
 Eq. \eqref{eq:general_matching_criterion} guarantees recognition success 
 by giving a lower bound on the number of 
 allowed erroneous pixels $\nfalse$.
 If a network stores a large number of patterns $M$, the minimal size 
 of the basins of attraction 
will be quite small and few erroneous pixels $\nfalse$ can be guaranteed 
to be corrected. In many applications, however, the number of patterns $M$ 
is much smaller than $N$ and the ability to correct larger errors is 
desired.

The last aspects to be discussed concern recognition time and
 oscillator accuracy. In physically all-to-all connected networks 
 (I) and (III), oscillator frequencies are not restrictive, as long as 
  they are similar enough to be well above the
   Kuramoto transition. 
  In contrast, frequency conditions in  dynamically all-to-all connected networks 
  limit the network size:
 Since in practice there will be only a certain frequency interval available,
 the number of oscillators is limited by the accuracy of the frequencies \cite{Hoelzel2011}.
 Recognition times have not been calculated analytically for any 
 of the oscillatory neural networks presented here. However, we assume that the frequency restrictions 
 present in (IIA), (IIB) and MONACO lead to slower recognition times compared to 
 (I) or (III). 
 Nevertheless, the shift of frequency due to the acceleration effect 
 \cite{Aonishi2002_AccelerationEffect} present in the real dynamics
 of  (IIA) and (IIB) \cite{Hoppensteadt1999,Hoelzel2011,Hoelzel2015_stability} 
  does not interfere with the recognition
  process in MONACO since the change in frequency is identical 
  in each oscillator pair due to its mirrored structure.
 Additionally, it is likely that the introduction of several coupling modulations 
 per subnetwork similar to the transition from architecture (IIA) to (IIB) 
  is possible for the MONACO-architecture. 
 In this improved network the scaling of necessary frequencies 
 would be reduced to $\mathcal{O}(N^{\ln(3)/ \ln(2)})$. 
\section{Summary \& Outlook}
\label{sec:Summary+Outlook}
We presented a network of coupled nonlinear oscillators as a 
 new architecture for an autoassociative memory device. 
Two subnetworks of oscillators with equal frequency distributions are each
globally coupled. An additional temporal modulation of the coupling  
is constructed from  
signals of oscillators in the other subnetwork 
 and the binary memorized patterns. 
The oscillator pairs of equal frequency synchronize to 
phase shifts of either $0$ or $\pi$, which corresponds to the 
pixel values of a binary output pattern.
Furthermore, the number of connections scales linearly 
with the number of pixels $N$ and the only necessary input are defective 
pattern and memorized patterns. 
 While 
orthogonal memorized patterns are always attractors, 
general memorized patterns
are stable as well if their projections on each other 
are not too large. 
Although spurious attractors also exist,  
 we derived a simple criterion for guaranteed recognition from 
worst case approximations on the basins of attraction for orthogonal 
as well as general patterns. 
Finally, our results were confirmed by simulations, which also indicate  
 that failed recognitions might occur but are quite rare 
as long as the criterion for guaranteed recognition is only weakly missed.
While other oscillatory neural networks exist which scale linearly with 
the number of pixels $N$ or have isolated attractors, our 
MONACO-architecture 
is the first to combine both features as well as the first to provide 
solid criteria for guaranteed recognition.

Several questions remain open:
The first concerns the maximal possible value of $\epsilon$ since a 
large $\epsilon$ decreases recognition time. 
Additionally, the system's robustness to frequency deviations or 
noise has not been quantified yet and might be addressed  
 both theoretically or experimentally. In principle, 
 a hardware realization is independent of the exact type of oscillator  
 as long as their signal shape is close to harmonic. Thus, also fast 
 state-of-the-art nano-oscillators \cite{Csaba_2013_SpinTorque_network,Jackson2015_Nano-Oscillators_RRAM_Cells} are conceivable. 
In this context, 
the influence of small delays should be discussed.

Finally, an even better time-connection tradeoff might be possible: 
Distributing frequency components on multiple 
{\cf{}}s similar to \cite{Kostorz2013} might provide 
better scaling of recognition time combined with all the other benefits 
of our architecture.
\begin{acknowledgments}
We want to thank Alexander Sparber and Stefan Litzel for fruitful 
discussions and careful reading. Furthermore, we thank the 
Nanosystems Initiative Munich (NIM) for 
 financial support.
\end{acknowledgments}
\appendix
\section{Averaging and frequency restrictions}
\label{app:averaging}
In this Appendix, we apply the method of averaging \cite{verhulst1996nonlinear} to the phase 
description Eq. \ref{eq:phase_equations}.
Therefore, we first expand the products in Eq. \eqref{eq:phase_equations} with the trigonometric equalities 
$\sin{x}\cos{y}=[\sin{(x-y)}+\sin{(x+y)}]/2$ and 
$\sin{x}\sin{y}=[\cos{(x-y)}-\cos{(x+y)}]/2$ to obtain all 
frequency components:
\setlength{\arraycolsep}{1pt}
\renewcommand{\arraystretch}{1.3}
\begin{equation*}
\begin{array}{lrcl}
\dot{\vartheta}^{[1]}_i -\Omega_i=&&&\\
\multicolumn{4}{l}{=\frac{\epsilon}{N}\cos{\vartheta^{[1]}_i} \cdot
	\sum_{j=1}^N\limits a^{[2]}(t)\sin{\vartheta^{[1]}_j}  \myvspace } \\
\multicolumn{4}{l}{=\frac{\epsilon}{N} \sum_{j,k,l=1}^N\limits \cmatrix_{lk} 
	\sin{\vartheta^{[2]}_l}\cos{\vartheta^{[1]}_i}\cdot  
	\sin{\vartheta^{[1]}_j}\sin{\vartheta^{[2]}_k} \myvspace}\\
= \frac{\epsilon}{4N} \sum_{j,k,l=1}^N\limits \cmatrix_{lk}& 
	\Big [&\multicolumn{2}{l}{\sinAv{\vartheta^{[2]}_l-\vartheta^{[1]}_i} + \sinAv{\vartheta^{[2]}_l+\vartheta^{[1]}_i}  \Big ]}\\
	\multicolumn{2}{r}{\cdot\Big [ }&\multicolumn{2}{l}{\cosAv{\vartheta^{[1]}_j-\vartheta^{[2]}_k} - \cosAv{\vartheta^{[1]}_j+\vartheta^{[2]}_k}  \Big ]  \myvspace}  \\
= \frac{\epsilon}{8N} \sum_{j,k,l=1}^N\limits \cmatrix_{lk} &
\Big [ & &\sinAv{\vartheta^{[2]}_l-\vartheta^{[1]}_i -\vartheta^{[1]}_j+\vartheta^{[2]}_k } \vspace{-7pt}\\
	&&+&\sinAv{\vartheta^{[2]}_l-\vartheta^{[1]}_i +\vartheta^{[1]}_j-\vartheta^{[2]}_k } \\
	&&-&\sinAv{\vartheta^{[2]}_l-\vartheta^{[1]}_i -\vartheta^{[1]}_j-\vartheta^{[2]}_k } \\
	&&-&\sinAv{\vartheta^{[2]}_l-\vartheta^{[1]}_i +\vartheta^{[1]}_j+\vartheta^{[2]}_k } \\
	&&+&\sinAv{\vartheta^{[2]}_l+\vartheta^{[1]}_i -\vartheta^{[1]}_j+\vartheta^{[2]}_k } \\
	&&+&\sinAv{\vartheta^{[2]}_l+\vartheta^{[1]}_i +\vartheta^{[1]}_j-\vartheta^{[2]}_k } \\
	&&-&\sinAv{\vartheta^{[2]}_l+\vartheta^{[1]}_i -\vartheta^{[1]}_j-\vartheta^{[2]}_k } \\
	&&-&\sinAv{\vartheta^{[2]}_l+\vartheta^{[1]}_i +\vartheta^{[1]}_j+\vartheta^{[2]}_k }  \Big ]
\end{array}
\end{equation*}

As $\vartheta^{[1/2]}_i=\Omega_i t +\mathcal{O}(\epsilon)$, each of the 
$\sin$-terms might oscillate with frequencies of $\mathcal{O}(\Omega_i)$
 or $\mathcal{O}( \Delta\Omega_{ij})$.($\Delta\Omega_{ij}=\Omega_i-\Omega_j$)
 As the characteristic timescales $\mathcal{O}({\Omega_i}^{-1})$ 
 and $\mathcal{O}({\Delta\Omega_{ij}}^{-1})$ are much smaller than 
 ${\epsilon}^{-1}$, the time average of these oscillating terms vanishes on times  
 $\mathcal{O}({\epsilon}^{-1}) \gg \mathcal{O}({\Delta\Omega_{ij}}^{-1})$.
If frequencies in the argument cancel each other out, however, the 
argument is constant on timescales 
 $\mathcal{O}({\epsilon}^{-1})$ and all oscillating terms are negligible. 
 Depending on the signs in the $\sin$-argument, there can be different 
 possibilities how constant terms can arise:
 
 
 In the first term, for example, frequencies cancel if
  $\Omega_l+\Omega_k = \Omega_i + \Omega_j$. That is always true for 
  $l=i \land k=j$ or $l=j \land k=i$, imposing an interaction between 
  the $i$-th and $j$-th oscillators in both networks depending 
  on $\cmatrix_{ij}$. However, frequencies might also cancel if the 
  frequency distribution is chosen poorly, which would wrongly connect 
  oscillators with different numbers $i,j,k,l$ only dependent 
  on $\cmatrix_{lk}$. Therefore, we require 
  $\Omega_l+\Omega_k \neq \Omega_m + \Omega_n 
  \,\forall\text{ pairwise different }l,k,m,n \label{eq:freq_condition++--}$.
  
  Similarly, the lowest order is obtained in the third term for 
  $\Omega_l=\Omega_i + \Omega_j + \Omega_k$. In order to avoid 
   interaction between the $i$-th and $j$-th oscillators based on 
   $\cmatrix_{lk}$ again, the frequency distribution must obey 
   $\Omega_l\neq \Omega_m + \Omega_n + \Omega_k  \,\forall l,k,m,n $ 
   and the third term becomes negligible as well as the fourth, fifth 
   and sixth term.
   
   While the eighth term averages out without further conditions, we 
   get identical contributions from the second and the seventh term. This 
   can be seen by renaming indices l and k and using $\cmatrix_{lk}=\cmatrix_{kl}$:

\begin{align*}
&\dot{\vartheta}^{[1]}_i -\Omega_i\approx\\
\approx&\frac{\epsilon}{8N} \sum_{j,k,l=1}^N\limits 
	\cmatrix_{lk} \Big [   \\
	&(\delta_{il}\delta_{kj} + \delta_{ik}\delta_{lj} - \delta_{ik}\delta_{kl}\delta_{lj})
		\sinAv{\vartheta^{[2]}_l-\vartheta^{[1]}_i -\vartheta^{[1]}_j+\vartheta^{[2]}_k }  \\
&+2 \delta_{il}\delta_{kj}
	\sinAv{\vartheta^{[2]}_l-\vartheta^{[1]}_i +\vartheta^{[1]}_j-\vartheta^{[2]}_k } \Big ]\\
 =&\frac{\epsilon}{8N} \sum_{j=1}^N\limits \Big [
	 -\cmatrix_{ij} 
	\sin \biglb (\big ( \vartheta^{[1]}_i -\vartheta^{[2]}_i \big ) 
		+ \big ( \vartheta^{[1]}_j -\vartheta^{[2]}_j \big ) \bigrb )  \\
 &- \cmatrix_{ji} \sin\biglb (\big ( \vartheta^{[1]}_i -\vartheta^{[2]}_i \big ) + \big ( \vartheta^{[1]}_j -\vartheta^{[2]}_j \big ) \bigrb ) 
		   \\
 &+2 \cmatrix_{ij}
	\sin\biglb (\big ( \vartheta^{[1]}_i -\vartheta^{[2]}_i \big ) - \big ( \vartheta^{[1]}_j -\vartheta^{[2]}_j \big )\bigrb ) \Big ] \\
	& + \frac{\epsilon}{8N}\cmatrix_{ii} 
	\sin \biglb (2\big ( \vartheta^{[1]}_i -\vartheta^{[2]}_i \big ) \bigrb ) \\
\end{align*}
Final simplifications can be obtained by introducing the phase difference of 
oscillators with identical frequency $\deltheta_i=\vartheta^{[1]}_i -\vartheta^{[2]}_i$ 
and using  $\cmatrix_{ji}=\cmatrix_{ij}$ as well as $S_{ii}=M$.
For $\dot{\vartheta}^{[2]}_i$, the calculation is the same with inverted upper indices:
\begin{align*}
	\dot{\vartheta}^{[1]}_i 
		&=\Omega_i+ \frac{\epsilon M}{8N}\sinAv{2\deltheta_i}\\
		&-\frac{\epsilon}{4N}\sum_{j=1}^N\limits\cmatrix_{ij}   
			\left [ \sinAv{\deltheta_i+\deltheta_j} + \sinAv{\deltheta_i-\deltheta_j} \right ]\\
	\dot{\vartheta}^{[2]}_i 
		&=\Omega_i- \frac{\epsilon M}{8N}\sinAv{2\deltheta_i}\\
		&+\frac{\epsilon}{4N}\sum_{j=1}^N\limits\cmatrix_{ij} 
		\left [ \sinAv{\deltheta_i+\deltheta_j} + \sinAv{\deltheta_i-\deltheta_j} \right ] \\
\end{align*}

{\bf Remark:}\\
As shown in \cite{Hoelzel2011}, both conditions on the frequency distribution can be 
simplified further:($l,k,m,n$ are still pairwise different.)
\begin{align*}
	\Omega_l+\Omega_k &\neq \Omega_m + \Omega_n \\
  \Omega_l-\Omega_n &\neq \Omega_m - \Omega_k \\
  \Delta\Omega_{ln} &\neq \Delta\Omega_{mk} \label{eq:freq_condition++--}\\
\end{align*}
All difference frequencies have to be different to each other. This can 
be fulfilled by multiplying the minimal difference frequency 
$\Delta\Omega_{min}$ with a Golomb-ruler \cite{golomb1997}, a set of integers with non-equal 
differences.
Similarly, the second condition can be simplified to 
$\Omega_l-\Omega_m=\Delta\Omega_{lm}\neq \Omega_n+\Omega_k\,\forall l,k,m,n$. 
 This last inequality is always fulfilled if  $\Omega_{min} > \Omega_{max}/3$.
\section{Ljapunov function and unstable fixed point sets}
\label{app:potential}
In this Appendix, we derive a Ljapunov function for Eq. \eqref{eq:theta_dynamics}. 
We use it to show that all fixed points with at least one index $i$ that fulfills Eq. \ref{eq:def_unstableSet} are unstable.  
%
\subsection{Ljapunov function}
First, we express Eq. \eqref{eq:theta_dynamics} as a gradient system 
with potential $U$, 
where $\Delta\dot{\vartheta}_i=-\partial U/\partial\deltheta_i \,\forall i$:
\begin{equation*}
	 U=-\frac{\epsilon}{2N} \sum_{l=1}^N\limits\Bigg (  
	 \sum_{k=1}^N\limits \cmatrix_{kl}\cos{\deltheta_k}\cos{\deltheta_l} - \frac{M}{2}\cos^2{\deltheta_l}
	 \Bigg )
\end{equation*}
This is equivalent to the overdamped motion of a particle in an 
energy landscape, where $\dot{\vec{v}}\propto -\vec{\nabla} E$.
 Therefore, $U$ decreases along trajectories and is a Ljapunov-function, 
 which ensures that fixed points are the only attractors possible in Eq. \eqref{eq:theta_dynamics}.

\subsection{unstable fixed points}
In order to prove that all fixed points with some $i \in \unstableSet$ are unstable,
 we express the system state in pattern coordinates 
 $\vec{\sysvec}$ with $\sysvec_i=\cos{\deltheta_i}$ 
and insert the coupling matrix $\cmatrix_{ij}=\sum_{m=1}^M \memvec_i^m\memvec_j^m$ 
into our potential function $U$:
\begin{align*}
	U
&=-\frac{\epsilon}{2N} \sum_{l=1}^N\limits\Bigg (  \sum_{k=1}^N\limits \cmatrix_{kl}\cos{\deltheta_k}\cos{\deltheta_l} - \frac{M}{2}\cos^2{\deltheta_l}\Bigg ) \\
&=-\frac{\epsilon}{2N}\sum_{k,l=1}^N\Bigg (\sum_{m=1}^M \memvec_k^m\memvec_l^m \sysvec_k\sysvec_l  - \frac{M}{2}\delta_{kl}\sysvec_l^2\Bigg ) \\
&= -\frac{\epsilon}{2N}\Bigg ( \sum_{m=1}^M \left < \vec{\memvec}^m,\vec{\sysvec} \right > ^2 - \frac{M}{2}\left < \vec{\sysvec},\vec{\sysvec} \right >\Bigg )
\end{align*} 
Now consider a small perturbation $\gamma\hat{e}_i$ 
from a fixed point $\fpvec$ where $i \in \unstableSet$:
\begin{align*}
	&U(\fpvec+\gamma\hat{e}_i)=\\
	&=-\frac{\epsilon}{2N}\Bigg [ \sum_{m=1}^M 
	\big ( \left < \vec{\memvec}^m,\fpvec \right > +
	\underbrace{\left < \vec{\memvec}^m,\gamma\hat{e}_i \right >}_{=\gamma\memvec_i^m} \big ) ^2 
	- \frac{M}{2}\big (  \left < \fpvec,\fpvec \right > \\
	& \quad + 2 \left < \fpvec,\gamma\hat{e}_i \right > + \left < \gamma\hat{e}_i,\gamma\hat{e}_i \right >\big )\Bigg ]\\
	&=U(\fpvec) -\frac{\epsilon\gamma}{N}\underbrace{\Bigg ( \sum_{m=1}^M 
	 \memvec_i^m\left < \vec{\memvec}^m,\fpvec \right > - \frac{M}{2}\fpvecindex_i
	\Bigg )}_{=0\text{, as }i \in \unstableSet.\text{(see Eq. \eqref{eq:def_unstableSet})}} \\
	&\quad -\frac{\epsilon}{2N}\Bigg ( \sum_{m=1}^M\gamma^2 \underbrace{(\memvec_i^m)^2}_{=+1}  - \frac{M}{2}\gamma^2 \Bigg )\\
	&=U(\fpvec) -\frac{\epsilon M}{4 N}\gamma^2 \\
\end{align*}
As $U$ decreases close to $\fpvec$, there must be an unstable eigendirection
 and the fixed point must be unstable if at
  least one $i$ with $\sum_{j=1}^N\cmatrix_{ij}\cos{\deltheta_j^*}-\frac{M}{2}\cos{\deltheta_i^*}= 0$ exists.(i.e. $i\in \unstableSet$)

Therefore, only the isolated fixed points with $\sin{\deltheta_i^*} =  0\,\forall i$
 and $\cmatrix_{ik}\sin{\phasediff_l^*} - M/2\delta_{ik}\sin{\deltheta_i^*}\neq 0\,\forall i$ 
 can be attractors.\\
 $\,$ \\
\bibliography{./PRX_2016_citations}
\end{document}